\documentclass{aa} 

\usepackage{graphicx}
\usepackage{txfonts}
\usepackage{amsmath}
\usepackage{caption}
\usepackage{subcaption}
\usepackage[colorlinks=true, linkcolor=blue, urlcolor=blue, citecolor=blue,]{hyperref}
\bibpunct{(}{)}{;}{a}{}{,} % to follow the A&A style
\begin{document}

   \title{Impact of an inhomogeneous density distribution on selected observational characteristics of circumstellar disks}
   \titlerunning{Impact of an inhomogeneous density distribution on observational characteristics of circumstellar disks}

   \author{R. Brauer
          \and
          S. Wolf
          }

   \institute{University of Kiel, Institute of Theoretical Physics and Astrophysics,
              Leibnizstrasse 15, 24118 Kiel, Germany\\
              \email{rbrauer@astrophysik.uni-kiel.de \quad wolf@astrophysik.uni-kiel.de}
	      }

%   \date{Received September 15, 1996; accepted March 16, 1997}
 
  \abstract
   {Analysis of observations of circumstellar disks around young stellar objects is often based on disk models with smooth and continuous density distribution. However, spatially resolved observations with increasing angular resolution and dynamical models indicate that circumstellar disks are highly structured.}
   {We investigate the influence of different clumpy density distributions on selected physical properties and on the observable characteristics of circumstellar disks. In particular, these are the temperature distribution, the spectral energy distribution, the radial brightness profile and the degree of polarization of scattered stellar radiation.}
   {Based on radiative transfer modeling we calculated the temperature structure of the disk and simulate observational quantities in the thermal re-emission and scattering regime. The clumpy density distributions are realized using a two-phase medium approach with phases for the clumps and the medium in between. We compared our results to those obtained for a smooth and continuous density distribution to quantify the influence of clumps on internal physical parameters and observable quantities of circumstellar disks.}
   {Within the considered model space, the clumpiness has a significant impact on the disk temperature distribution. For instance, in the transition region from the optically thin upper disk layers to the disk interior, it causes a decrease in the mean temperature by up to $12~\mathrm{K}$ (corresponding to $\sim 15~\%$), if compared to continuous disks. In addition, circumstellar disks with clumpy density distributions generally feature a lower spectral index in the submm/mm range of the SED than continuous disks. The strength of this decrease can be varied by changing the dust mass or grain size, but not by changing the inclination of the disk. As a consequence of the lower spectral index, the dust grain size derived from the submm/mm-slope of the SED may be overestimated, if the inhomogeneity of the disk density distribution is not taken into account. Furthermore, the scattered light brightness distribution of clumpy disks shows a steeper radial decrease than in the case of continuous disks. The azimuthal variations in the scattered flux, resulting from inhomogeneous density distributions, have their maximum at the medium radial extent of the disks.
   Additionally, clumpy density distributions change the degree of polarization of the scattered light in the optical compared to continuous disks. The quantitative level of this variation increases with the optical depth of the clumps.}
   {}
% The results show that the structure of circumstellar disks are an important quality and must be considered to obtain adequate results. 

   \maketitle

%%%%%%%%%%%%%%%%%%%%%%%%%%%%%%%%%%%%%%%%%%%%%%%%%%%%%%%%
\section{Introduction}\label{intro}
%%%%%%%%%%%%%%%%%%%%%%%%%%%%%%%%%%%%%%%%%%%%%%%%%%%%%%%%
   During the past decade, light curve measurements and observations of large scale structures down to $100~\mathrm{au}$ reveal that circumstellar disks are rather highly structured than continuous (\citealt{2015arXiv150403568S}; \citealt{2014A&A...568A..40G}; \citealt{2014ApJ...781...87A}; \citealt{2013A&A...560A.105G}).
   In addition, hydrodynamic simulations show that effects such as vortices increase the local density of circumstellar disks by a factor of about 10 \citep{1999ApJ...523..350G, 2001MNRAS.323..601D}. An increase by the factor of 100 and even more is predicted in selected cases \citep{2000ApJ...537..396G, 2001MNRAS.323..601D, 2014arXiv1405.2790Z}. However, it is unclear how the density structure (clumpiness) influences parameters that are estimated from observations of circumstellar disks. Therefore, modeling and fitting of circumstellar disks are usually performed with a smooth and continuous density distribution \citep{2011epsc.conf.1031G, 2011A&A...527A..27S, 2012A&A...543A..81M}. Currently and in the near future, observatories and instruments operating in the infrared to millimeter wavelength regime will become available and will provide angular resolutions of up to a few mas (e.g., SPHERE/VLT, MATISSE/VLTI, ALMA). With these observatories and instruments, it will be possible to resolve circumstellar disks down to few au which will lead to better understanding the structure of circumstellar disks. However, the impact of a complex disk structure on the analysis of existing spatially unresolved or low-resolution observation has hardly been investigated so far.

   Previous investigations of the influence of inhomogeneous density distributions in different environments, excluding circumstellar disks, used various approaches to realize clumpiness \citep{1972ApJ...176..651M, 1984ApJ...287..228N, 1993MNRAS.264..161H, 1999ApJ...523..265V, 2000ApJ...528..799W}. For instance, \cite{1996ApJ...463..681W} and \cite{1998A&A...340..103W} used a spherical density distribution and a two-phase medium to implement clumps. Other approaches are n-phase media with $n>2$ \citep{1993MNRAS.264..145H}, massive spheres with radii depending on their positions \citep{2008A&A...482...67S} and stochastic density variations \citep{2003MNRAS.342..453H}. 

   In this study, we pursue two aims. On the one hand, we investigate how inhomogeneous density distributions change observational characteristics of circumstellar disks and their impact on estimations of selected internal physical and observational quantities. On the other hand, we investigate whether the clumpiness of circumstellar disks can be inferred from their physical and observable quantities. These characteristics are the spectral index in the submm/mm of the SED, the radial brightness profile and the degree of polarization of scattered stellar radiation. These quantities are derived through radiative transfer modeling of thermal re-emission and scattering of stellar radiation by the dust. We use a two-phase medium approach in spherical coordinates and the non-constant density distribution of circumstellar disks.

   This paper is structured as follows. We introduce our model in Sect. \ref{model}, including the disk model and the implementation of the two-phase medium, the dust properties, the heating source and our radiative transfer code. Subsequently, we present our results for the impact of clumpiness on the temperature distribution (Sect. \ref{temp_distr}), the spectral index (Sect. \ref{spectral_index}), the radial brightness profile (Sect. \ref{r_b_p}), and the degree of polarization (Sect. \ref{polarization}). Concluding discussions are compiled in Sects. \ref{discussion} and \ref{conclusions}.

%%%%%%%%%%%%%%%%%%%%%%%%%%%%%%%%%%%%%%%%%%%%%%%%%%%%%%%%
\section{Model description}\label{model}\label{disk}
%%%%%%%%%%%%%%%%%%%%%%%%%%%%%%%%%%%%%%%%%%%%%%%%%%%%%%%%

   \subsection*{Disk structure:}

   Our disk model is based on the density distribution from the work by \cite{1973A&A....24..337S} which can be written as
   \begin{equation}
    \rho_\mathrm{disk}=\rho_0 \left( \frac{R_\mathrm{ref}}{\overline{\omega}} \right)^\alpha \exp\left(-\frac{1}{2}\left[\frac{z}{h(\overline{\omega})}\right]^2 \right).\label{eq:disk}
   \end{equation}
   Here, $\overline{\omega}$ is the radial distance from the star in the disk midplane, $z$ is the distance from the midplane of the disk, $R_\text{ref}$ is a reference radius and $h(\overline{\omega})$ is the scale height. The density $\rho_0$ is chosen according to the preset dust mass. For clumpy disks, $\rho_0$ is adjusted to obtain the same disk mass as for the corresponding smooth reference disk. The scale height is a function of $\overline{\omega}$ as follows:
   \begin{equation}
    h(\overline{\omega})=h_0 \left(\frac{\overline{\omega}}{R_\mathrm{ref}}\right)^\beta.\label{eq:disk2}
   \end{equation}
   The parameters $\alpha$ and $\beta$ set the radial density profile and the disk flaring, respectively. The extent of the disk is constrained by the inner radius $R_\mathrm{in}$ and the outer radius $R_\mathrm{ou}$.

   We consider circumstellar disks with dust masses ranging from $10^{-6}~\mathrm{M_\odot}$ to $10^{-4}~\mathrm{M_\odot}$ (corresponding to total masses of the disks of $\sim 10^{-4}~\mathrm{M_\odot}\ldots10^{-2}~\mathrm{M_\odot}$). These lie in the lower range of class $2/3$ YSO disks typically \citep{2006ApJS..167..256R}.

   \label{clumpy_distr}
   While the underlying smooth and continuous density distribution is described by Eq. \ref{eq:disk}, we now introduce a two-phase medium approach to realizing clumpy density distributions. From now on, we denote the smooth and continuous density distribution as the ``reference disk'' or the ``reference density distribution''. The values of the parameters, which we introduce here and in the following sections, are summarized in Tabl. \ref{tab:parameter}. Our two-phase medium is motivated by the studies of \cite{1996ApJ...463..681W} and \cite{1998A&A...340..103W}. Here, each cell in the disk either represents the clump or the interclump phase (see Sect. \ref{mc3d}). The clumpy medium is characterized through the density contrast $k$ and a mass ratio $\mathit{\eta}$ between both phases. The latter is defined as the ratio between the total mass in the clumps ($M_\mathrm{clump}$) and the entire disk mass ($M_\mathrm{disk}$):
   \begin{equation}
    \mathit{\eta} = \frac{M_\mathrm{clump}}{M_\mathrm{disk}}.\label{eq:mass_ratio}
   \end{equation}

   Based on the reference density distribution (Eq. \ref{eq:disk}), the density of each cell depends on their position $(r,z)$, their related phase (clump, interclump) and the density contrast $k$ as follows:
   \begin{align}
    \rho_\mathrm{clump}(r,z) &= \rho_\mathrm{disk}(r,z)\cdot k,\\
    \rho_\mathrm{interclump}(r,z) &= \rho_\mathrm{disk}(r,z),\\
    k &= \frac{\rho_\mathrm{clump}(r,z)}{\rho_\mathrm{interclump}(r,z)}.
   \end{align}

   We use values of the density contrast $k$ that are consistent with hydrodynamic simulations (see Sect. \ref{intro} and Tabl. \ref{tab:parameter}). Values of the mass ratio $\eta$ are chosen to evenly cover the whole range from a few to many clumps (see Tabl. \ref{tab:parameter}).

\begin{figure}
    \centering
    \resizebox{0.9\hsize}{!}{\includegraphics[width=\hsize]{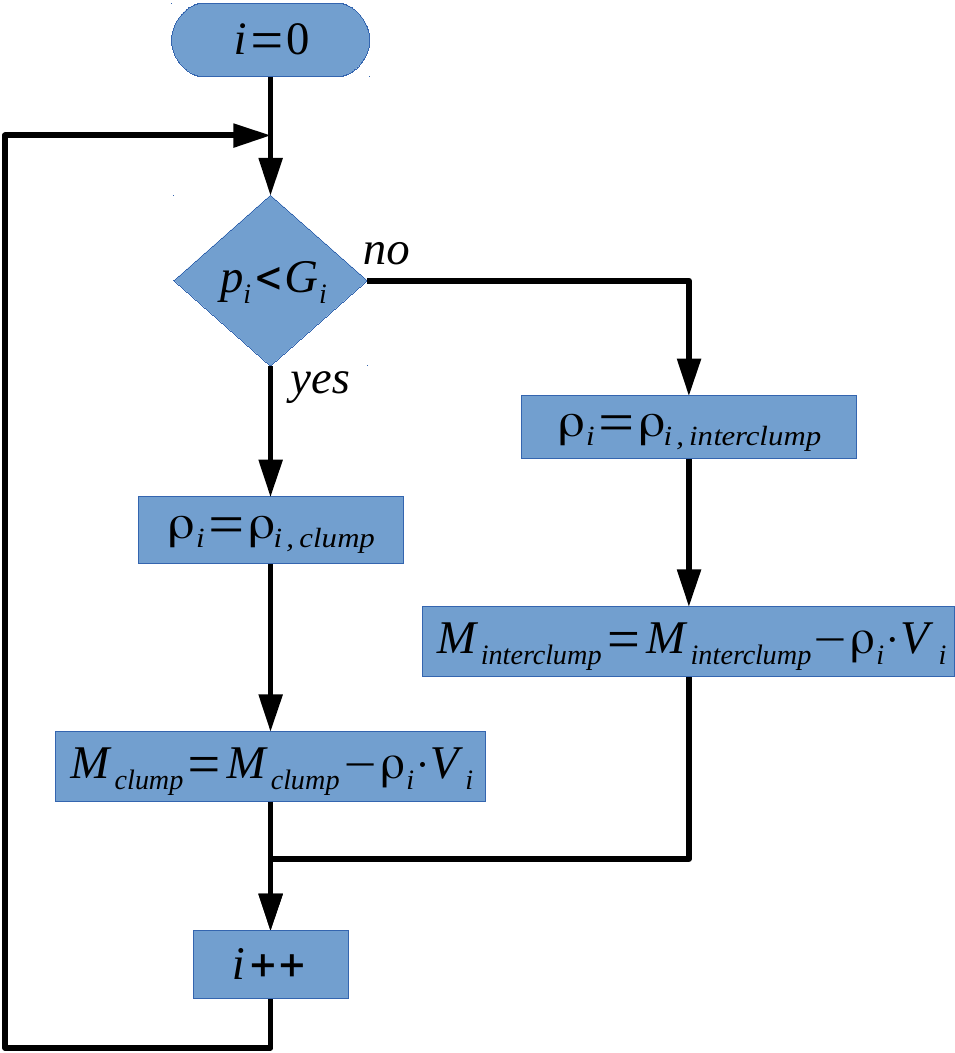}}
    \caption{Flowchart of the clump distribution algorithm. The quantity $i$ is the index number of a cell, while $V_i$ is the related cell volume. The quantity $G_\mathrm{i}$ describes the threshold which defines the probability that a cell represents the clump phase. The quantities $\rho_\mathrm{i,clump}$ and $\rho_\mathrm{i,interclump}$ describe the density of cell $i$, if it would be in the corresponding phase.}
    \label{fig:flowchart}
   \end{figure}

   The distribution of clumps is realized with a random number $p_\mathrm{i}\in[0,1)$ and a threshold $G_\mathrm{i}$ for each cell (index $i$). A cell is in the clump phase, if $p_\mathrm{i}$ is smaller than the threshold $G_\mathrm{i}$. The threshold of the first cell $G_\mathrm{0}$ is the mass ratio $\eta$. The threshold of a subsequent cell is calculated as follows:
   \begin{equation}
    G_\mathrm{i+1}=\frac{M_\mathrm{clump}}{M_\mathrm{interclump}+M_\mathrm{clump}},\label{eq:threshold}
   \end{equation}
   where $M_\mathrm{clump}$ and $M_\mathrm{interclump}$ are the respective masses which have to be distributed in the remaining cells to achieve the preset mass ratio $\eta$. A flowchart of this algorithm is displayed in Fig. \ref{fig:flowchart}.

   We use spherical coordinates with a step width factor in radial direction to implement the cells of our model (see Sect. \ref{mc3d}). Therefore, the length scales of the clump and interclump phase extend from a few $0.01~\mathrm{au}$ in the innermost regions to some $10~\mathrm{au}$ in the outermost regions.

\begin{figure*}
    \centering
    \resizebox{\hsize}{!}{\includegraphics[width=\hsize]{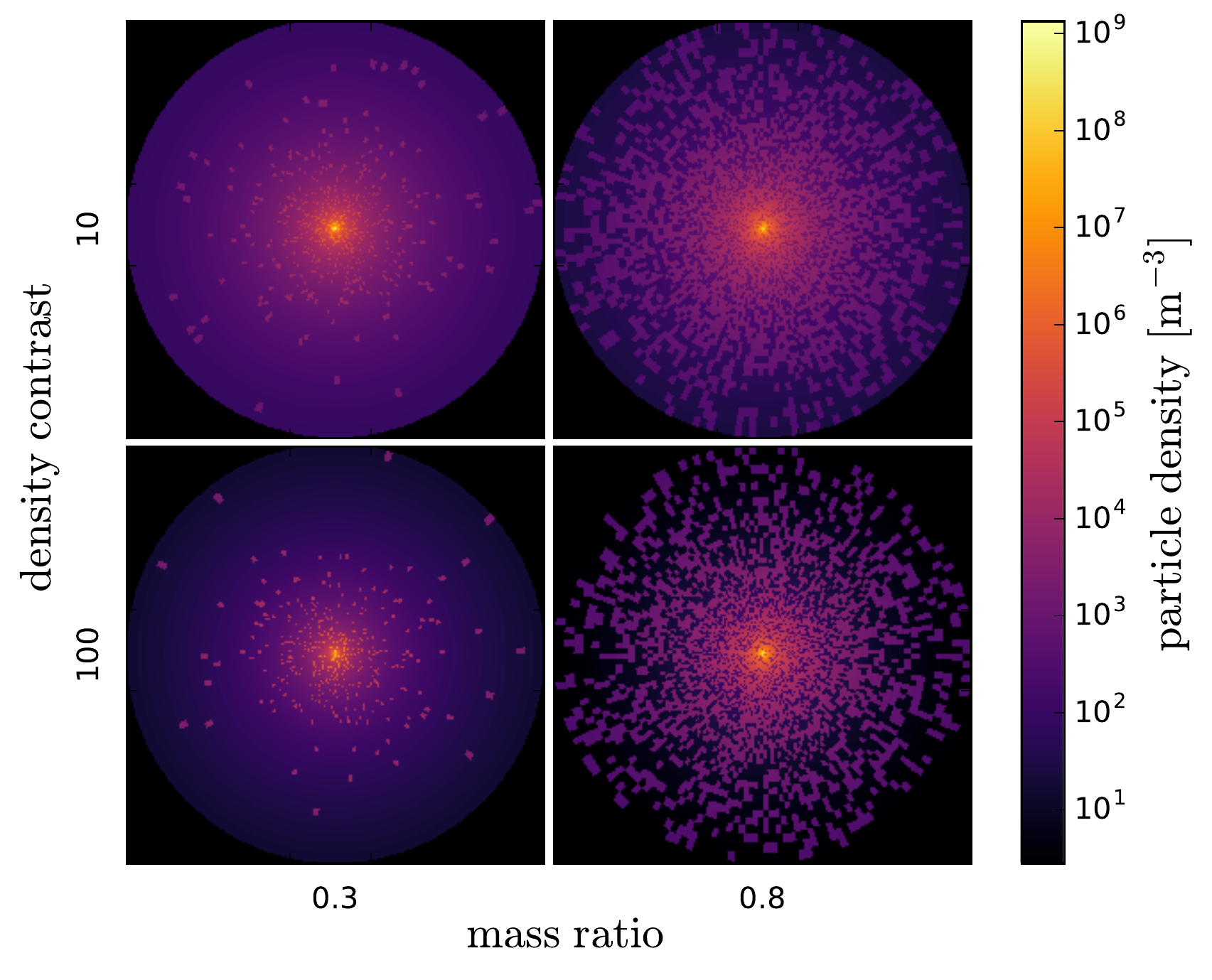}
    \includegraphics[width=\hsize]{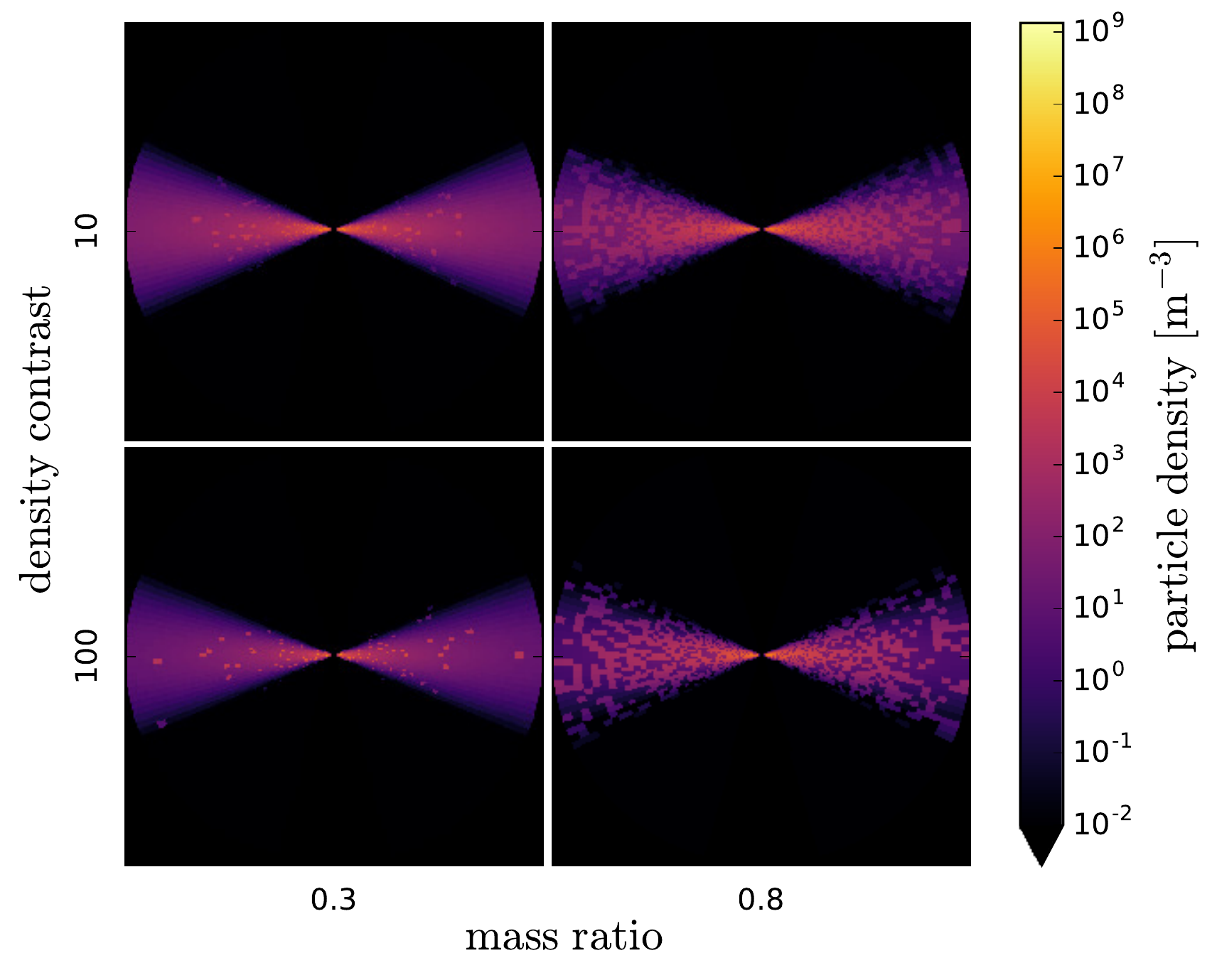}}
    \caption{Number density distribution in the disk midplane (left) and in a vertical cut through the disk (right) for four clumpy density distributions.}
    \label{fig:density_dist}
   \end{figure*}

   In Fig. \ref{fig:density_dist} the clumpy disk density distribution is illustrated for four selected combinations of the density contrast $k$ and mass ratio $\eta$.

\subsection*{Dust:}\label{dust}

   We assume compact, homogeneous and spherical grains, consisting of 62.5\% silicate and 37.5\% graphite. \citep[optical properties from][]{2001ApJ...548..296W}. For the grain size distribution we assume\vspace{-0.011cm}
   \begin{equation}
    \mathrm{d}n(a)\propto a^{-3.5} \text{d}a, \quad a_\mathrm{min} < a < a_\mathrm{max},
   \end{equation}
   where $n(a)$ is the number of dust particles with a specific dust grain radius $a$.

\subsection*{Heating source:}\label{heating}

   The primary heating source is a central pre-main-sequence star. The star is characterized by an effective temperature $T_\mathrm{star}$ and a stellar radius $R_\mathrm{star}$ which are summarized in Tabl. \ref{tab:parameter}.

\subsection*{Radiative transfer:}\label{mc3d}

   We apply the three-dimensional Monte Carlo continuum RT code MC3D in spherical coordinates \citep{1999A&A...349..839W,2003CoPhC.150...99W} to calculate the spatial temperature distribution of the disk and to derive the observable quantities discussed in Sect. \ref{results}. It solves the radiative transfer problem self-consistently with the Monte Carlo method. Thermal re-emission as well as scattering by dust are considered. The required optical properties of the dust grains are derived from the refractive index of the given material by using the MIEX scattering routine (see Sect. \ref{dust}; \citealt{2004CoPhC.162..113W}). The main assumption of MC3D and other RT codes is that the model is sufficiently approximated by cells with constant density and temperature.

   For the simulation of the scattered light, we apply the peel-off-technique \citep{1984ApJ...278..186Y, 2001ApJ...551..269G}. Its advantage is a high signal-to-noise ratio of scattered light maps which is particularly useful for dense, i.e. optically thick regions such as clumps.
   
{\renewcommand{\arraystretch}{1.2}
\begin{table}
\caption{Overview of model parameters.}
\label{tab:parameter}
\begin{tabular}{lcl}
  \hline
  \hline
 \multicolumn{3}{c}{\textbf{central star}} \\
  \hline
 effective temperature  &  $T_\mathrm{star}$ & $4000~\mathrm{K}$ \\
 stellar radius  &  $R_\mathrm{star}$ & $2~\mathrm{R_\odot}$ \\
 distance to star  &  $d$ & $140~\mathrm{pc}$ \\
 wavelengths & $\lambda$ & $[0.05, 2000]~\mathrm{\mu m}$\\
   \hline
  \multicolumn{3}{c}{\textbf{disk model}}\\
   \hline
 inner radius  &  $R_\mathrm{in}$ & $0.1~\mathrm{au}$ \\
 outer radius  &  $R_\mathrm{ou}$ & $300~\mathrm{au}$ \\
 scale height  &  $h_0$ & $10~\mathrm{au}$ \\
 characteristic radius  &  $R_\mathrm{ref}$ & $100~\mathrm{au}$ \\
 radial density decrease  &  $\alpha$ & $2.625$ \\
 disk flaring  &  $\beta$ & $1.125$ \\
 cells in $r$-direction  &  $n_r$ & $99$\\
 step width factor in $r$  &  $\mathrm{sf}$ & $1.05$\\
 cells in $\theta$-direction  &  $n_\theta$ & $91$\\
 cells in $\phi$-direction  &  $n_\phi$ & $180$ \\
 inclination & $i$ & $[0^\circ,60^\circ,90^\circ]$ \\
   \hline
  \multicolumn{3}{c}{\textbf{dust}} \\
   \hline
 dust grain density  &  $\rho_\mathrm{dust}$ & $2.5~\mathrm{g\ cm^{-3}}$\\
 minimum grain size  &  $a_\mathrm{min}$ & $0.005~\mathrm{\mu m}$ \\
 maximum grain size  &  $a_\mathrm{max}$ & $[0.25,10]~\mathrm{\mu m}$ \\
 dust mass  &  $M_\mathrm{dust}$ & $[10^{-6},10^{-5},10^{-4}]~\mathrm{M_\odot}$ \\
   \hline
   \multicolumn{3}{c}{\textbf{clumps}} \\
   \hline
 mass ratio  &  $\eta$ & $[0.3,0.5,0.7,0.8,0.95]$ \\
 density contrast  &  $k$ & $[3.2,10,32,100]$\\
   \hline
\end{tabular}
\end{table}
{\renewcommand{\arraystretch}{1}

%%%%%%%%%%%%%%%%%%%%%%%%%%%%%%%%%%%%%%%%%%%%%%%%%%%%%%%%
\section{Results}\label{results}
%%%%%%%%%%%%%%%%%%%%%%%%%%%%%%%%%%%%%%%%%%%%%%%%%%%%%%%%

   At first we discuss the impact of clumpiness on  the spatial temperature distribution in the disk (Sect. \ref{temp_distr}). Subsequently, the spectral index in the submm/mm range of the thermal dust re-emission is investigated (Sect. \ref{spectral_index}). In the second part of our study, we compare the radial brightness profiles and polarization maps of clumpy and continuous disks (Sect. \ref{r_b_p}, \ref{polarization})

%%%%%%%%%%%%%%%%%%%%%%%%%%%%%%%%%%%%%%%%%%%%%%%%%%%%%%%%
\subsection{Temperature distribution}\label{temp_distr}
%%%%%%%%%%%%%%%%%%%%%%%%%%%%%%%%%%%%%%%%%%%%%%%%%%%%%%%%

   \begin{figure}[t]
    \includegraphics[width=\hsize]{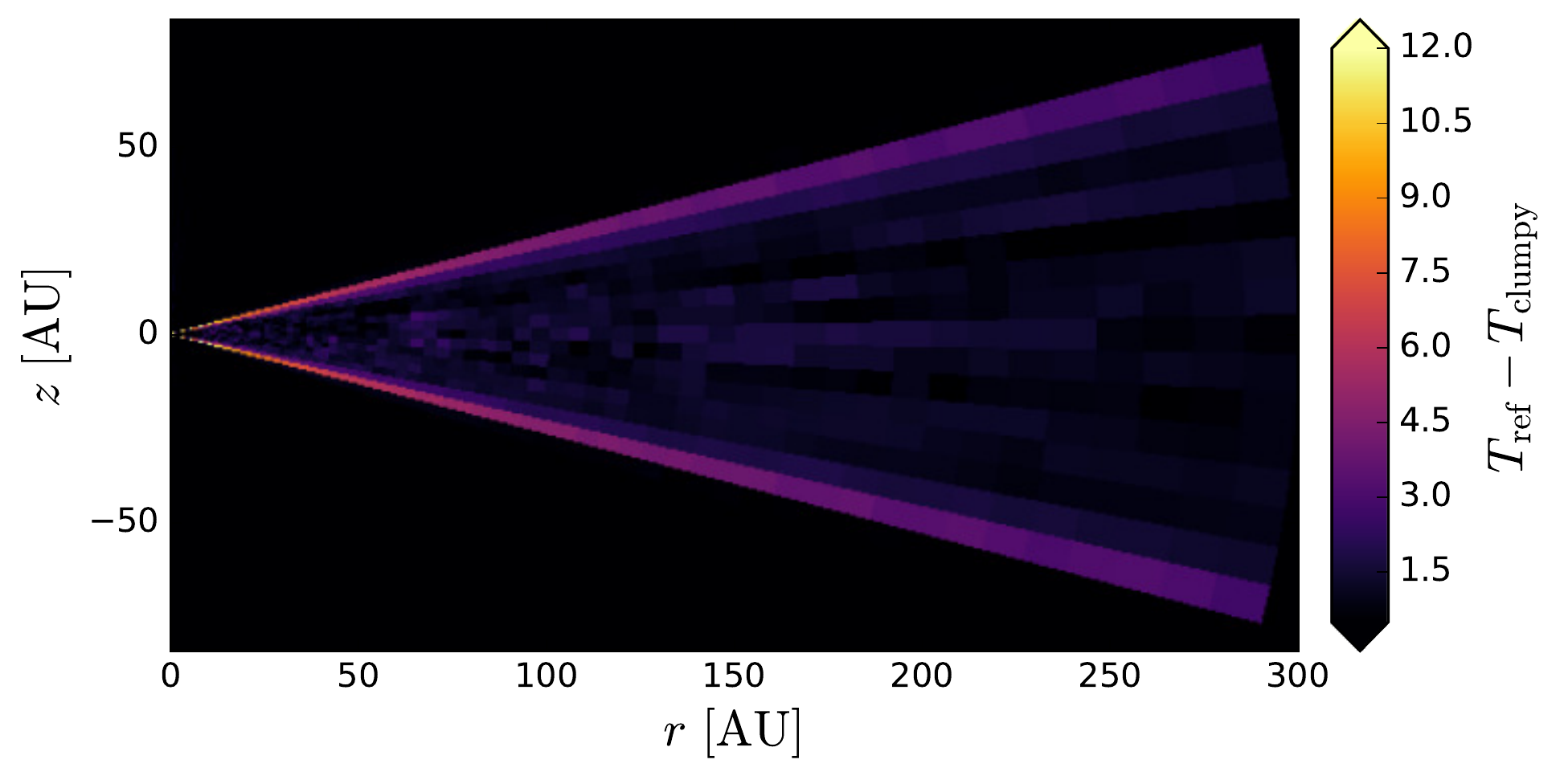} \\ \includegraphics[width=\hsize]{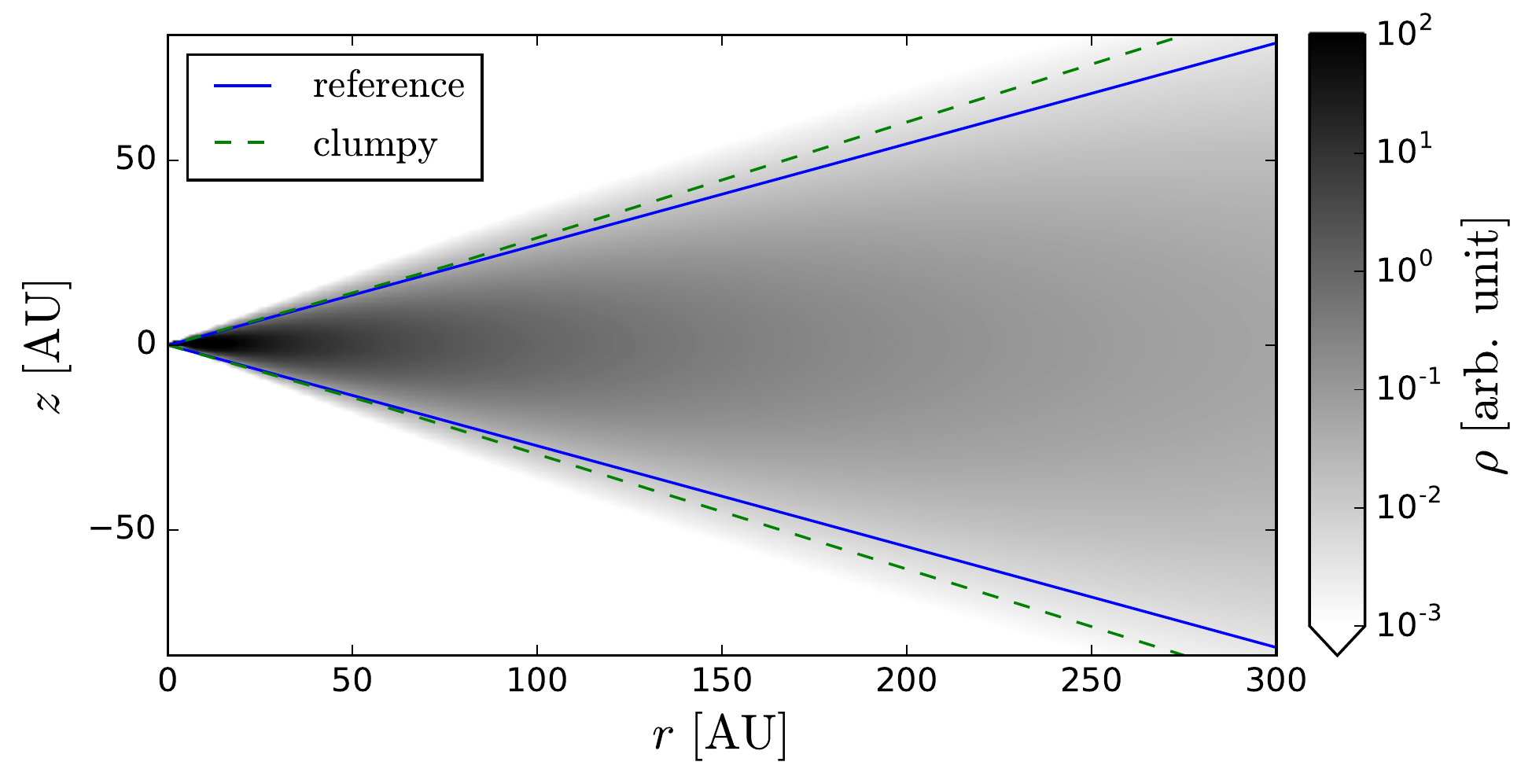}
    \caption{Temperature difference between the reference disk and a selected clumpy disk (top) and lines with optical depth $\tau\sim1$ of our reference disk and the clumpy disk (bottom). The density distribution in the bottom image is taken from Eq. \ref{eq:disk}. The optical depth is calculated at the wavelength of maximum stellar emissivity in the way of the stellar radiation. ($\eta=0.3$, $k=32$, $i=0^\circ$, $M_\mathrm{dust}=10^{-6}~\mathrm{M_\odot}$, $a_\mathrm{min}=0.005~\mathrm{\mu m}$ and $a_\mathrm{max}=0.25~\mathrm{\mu m}$)}
    \label{fig:temp_plane_diff}
   \end{figure}

   \begin{figure}[t]
    \includegraphics[width=\hsize]{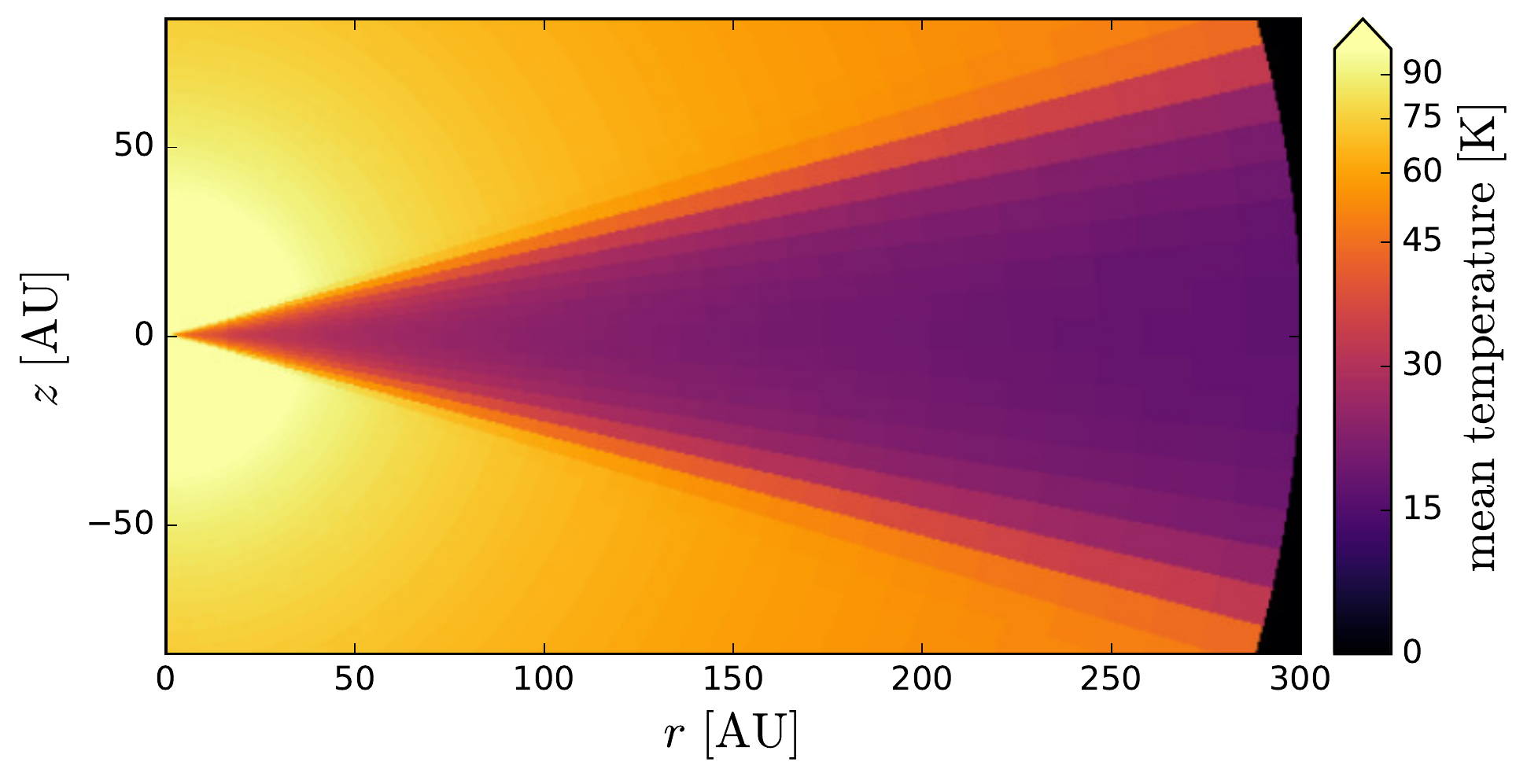} \\ \includegraphics[width=\hsize]{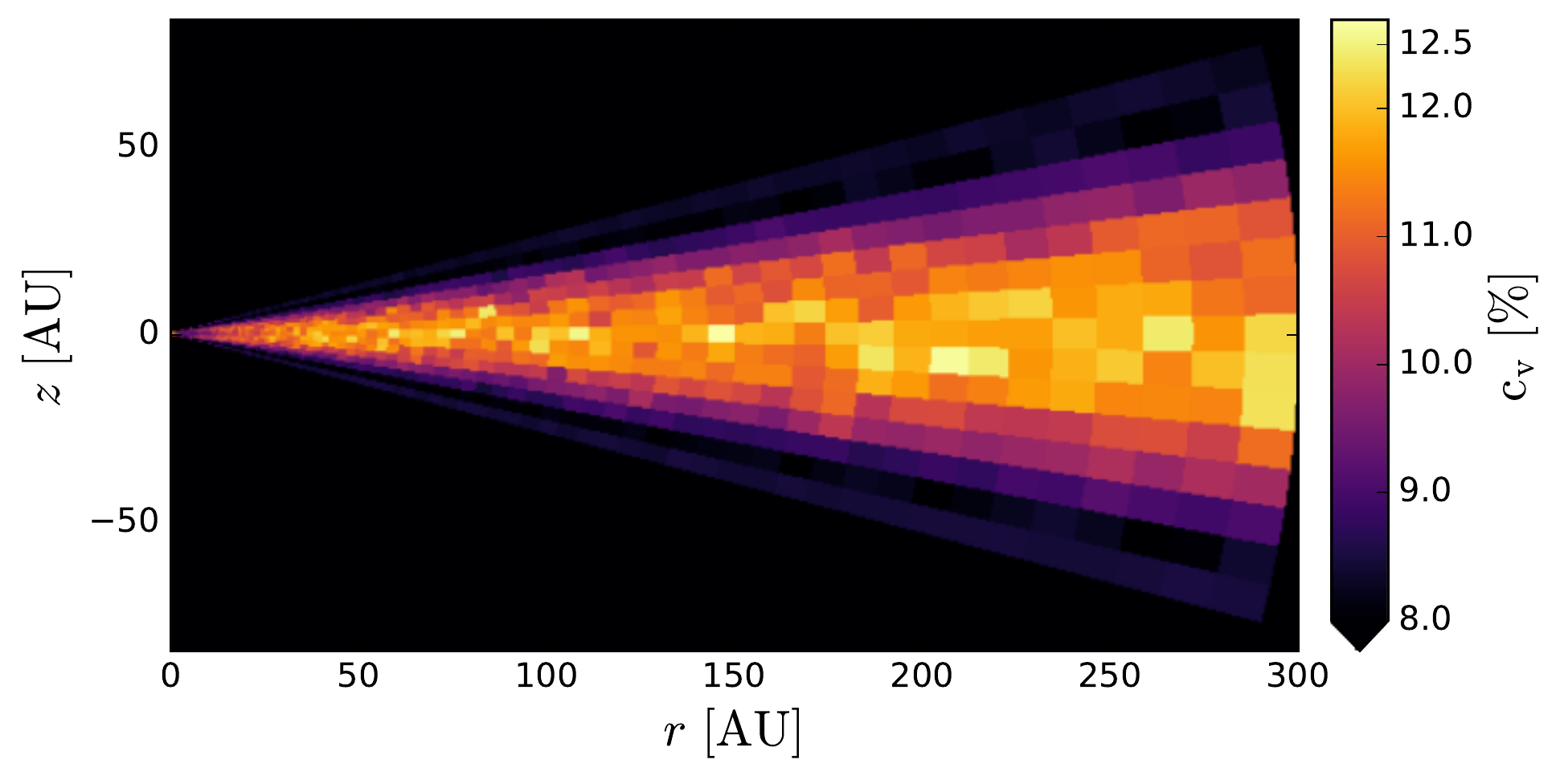}
    \caption{Vertical cut through the azimuthally averaged temperature distribution (top) and the related coefficient of variation (bottom). ($\eta=0.3$, $k=32$, $i=0^\circ$, $M_\mathrm{dust}=10^{-6}~\mathrm{M_\odot}$, $a_\mathrm{min}=0.005~\mathrm{\mu m}$ and $a_\mathrm{max}=0.25~\mathrm{\mu m}$)}
    \label{fig:temp_plane_clumpy}
   \end{figure}

   At first, we examine the impact of the clumpiness on the spatial temperature distribution. Therefore, we simulate the temperature distribution of a circumstellar disk with a dust mass of $10^{-6}~\mathrm{M_\odot}$ and a dust grain size distribution ranging from $0.005~\mathrm{\mu m}$ to $0.25~\mathrm{\mu m}$. We performed this for our reference disk and for a clumpy disk with $\mathit{\eta}=0.3$ and $k=32$. 

   To illustrate the quantitative differences in the temperature distribution between a clumpy and the reference disk, we first average the temperature distribution in azimuthal direction. Subsequently, we create a difference image (Fig. \ref{fig:temp_plane_diff}, top). The reference disk has a slightly higher temperature in the optically thick region close to the disk midplane ($\Delta T\sim3~\mathrm{K}$) and a significantly higher temperature in the transition region from the optically thin upper disk layers to the disk interior ($\Delta T\sim[5,12]~\mathrm{K}$). In the transition region, the optical depth is still low enough to allow direct heating of the dust by stellar radiation ($\tau\sim1$ at wavelength of maximum stellar emissivity). However, if the dust is distributed inhomogeneously in this region, previously still optically thin regions may become optically thick. Therefore, the heating efficiency of these regions decreases significantly. At the same time, clumps (if optically thick) cast shadows over the optically thin regions of the interclump medium in the transition region, not allowing the stellar radiation to efficiently heat them either. Consequently, the entire former $\tau\sim1$ region becomes cooler as compared to the reference disk. To confirm this, the density profile of the reference disk and the $\tau\sim1$ line at the wavelength of maximum stellar emissivity is shown in Fig. \ref{fig:temp_plane_diff} (bottom). For illustration of the explanation given above, the azimuthally averaged $\tau\sim1$ line of the clumpy disk is also shown this image.

   In addition, we examined the increase in the temperature inside the clumpy disk. The azimuthally averaged temperature distribution of the clumpy disk and the related coefficient of variation is displayed in Fig. \ref{fig:temp_plane_clumpy}. The coefficient of variation $c_v$ is calculated as follows \citep{R265}:
   \begin{equation}
    c_v = \frac{\sigma}{\mu}. 
   \end{equation}
   Here, $\sigma$ is the standard deviation and $\mu$ is the mean value of the dust temperature for a given radial distance and distance from the midplane. The stochastic distribution of clumps leads to a deviation in the temperature distribution as compared to the reference disk. The clumps near the midplane have significantly higher optical depths than the interclump-medium and are mainly optically thick. In contrast to the transition region, the dust near the midplane of the reference disk is also mainly optically thick. In a given direction, clumps enhance the extinction of radiation which results in lower temperatures in and behind them in comparison to the reference disk. In other directions, the interclump-medium causes the opposite effect and increases the temperature in comparison to the reference disk. For a given radial distance and distance from the midplane (quantified by the angle $\theta$), the temperature varies in azimuthal direction. Near the disk midplane this variation amounts to about $12\%$ of its mean value. In Fig. \ref{fig:temp_plane_clumpy} it is shown that the mean temperature exhibits only small variations in the disk interior ($T\sim[15,30]~\mathrm{K}$). As a result, the variations in the temperature in azimuthal direction are high enough to vary the radial distance of isothermal lines by several dozens of au.

   Additionally, we investigated the spatial temperature distributions of circumstellar disks with larger dust masses ($M_\mathrm{dust}\in[10^{-5},10^{-4}]~\mathrm{M_\odot}$). It shows that the temperature near the midplane decreases due to the higher optical depth. In contrast, the absolute azimuthal variations in the temperature remain almost constant in that region so that the coefficient of variation increases. In addition, the difference between the temperature in the transition region of the clumpy and the reference disk decreases in the outer parts of the disk by few degrees. The higher optical depth in the reference disk mainly causes the lower temperature difference by decreasing the temperature in the transition region. In addition, the distance of the transition region from the midplane increases by a few au.
 
%%%%%%%%%%%%%%%%%%%%%%%%%%%%%%%%%%%%%%%%%%%%%%%%%%%%%%%%
\subsection{Spectral index}\label{spectral_index}
%%%%%%%%%%%%%%%%%%%%%%%%%%%%%%%%%%%%%%%%%%%%%%%%%%%%%%%%

   The submm/mm-slope of a spectral energy distribution depends on the behavior of the absorption efficiency of the dust grains $Q_\mathrm{abs}(\lambda)$ and the behavior of the black body radiation in the Rayleigh-Jeans-limit $B_\mathrm{\lambda,RJ}(T)$. We characterize the submm/mm-slope with the spectral index $\alpha_\mathrm{mm}$ as follows \citep{2007prpl.conf..767N}:   
   \begin{align}
    B_\mathrm{\lambda,RJ}(T)&\propto\lambda^{-\alpha_\mathrm{1}},\quad \alpha_\mathrm{1}=2,\\
    Q_\mathrm{abs}(\lambda)&\propto\lambda^{-\alpha_\mathrm{2}},\\
    f_\lambda&\propto\lambda^{-(2+\alpha_\mathrm{2})}=\lambda^{-\alpha_\mathrm{mm}}.\label{eqn:spectral_index}
   \end{align}
   Here, the quantity $f_\lambda$ is the flux in the submm/mm wavelength range. Because of the Rayleigh-Jeans-limit, $\alpha_\mathrm{1}$ is constant. In contrast, the behavior of $\alpha_\mathrm{2}$ with wavelength depends on the dust grain size (distribution). Therefore, the submm/mm-slope is usually used to determine the size of the radiating dust grains. If $f_\lambda$ is measured in units of $[\mathrm{Jy}]$, the radiation of interstellar, submicron-sized dust grains results in a spectral index of $\alpha_\mathrm{mm}=4$.
   
   In an optically thin circumstellar disk, all dust grains contribute to the net flux and the spectral index is expected to only depend on the wavelength dependence of the absorption efficiency in the RJ-limit. However, even disks that are completely optically thin in the continuous case, can feature optically thick regions with inhomogeneous density distributions. In optically thick regions, not every dust grain contributes to the net flux. In addition, the decrease in the optical depth with wavelength shrinks optically thick regions and increases the number of dust grains that contribute to the net flux. Hence, the increase in the net flux flattens the submm/mm-slope which decreases $\alpha_\mathrm{mm}$. To take this into account, we add a third component to Eq. \ref{eqn:spectral_index} that depends on a new quantity $\zeta$. It describes the influence of the density distribution, i.e. optical depth, on the spectral index:
   \begin{align}
    \zeta&\propto\lambda^{-\alpha_\mathrm{3}},\quad \alpha_\mathrm{3}<0,\label{eqn:spectral_index_third1}\\
    f_\lambda&\propto\lambda^{-(2+\alpha_\mathrm{2}+\alpha_\mathrm{3})}=\lambda^{-\alpha_\mathrm{mm}},\quad \alpha_\mathrm{3}<0.\label{eqn:spectral_index_third}
   \end{align}

   It is expected that the absorption efficiency of a dust grain size distribution containing small grains is steeper with wavelength than a distribution containing larger grains. Therefore, the dust grain size will be overestimated, if $\zeta$ is not taken into account. Hence, we investigate the impact of clumpy density distributions defined by the parameters $k$ and $\eta$ on the spectral index in the submm/mm of the SED. In addition, we consider changes in the inclination $i$, the dust mass $M_\mathrm{dust}$ and the maximum dust grain size $a_\mathrm{max}$. For this purpose, we simulate a disk with a clumpy density distribution for any given combination of our model parameters ($k$, $\eta$, $i$, $M_\mathrm{dust}$, $a_\mathrm{max}$; see Tabl. \ref{tab:parameter} for parameter space). For comparison, we also simulate our reference disk ($k=1$, $\mathit{\eta}=1$) for every set of the parameters $i$, $M_\mathrm{dust}$ and $a_\mathrm{max}$. We calculate the spectral index by fitting Eq. \ref{eqn:spectral_index_third} on our SEDs. The fitting range extends from $763~\mathrm{\mu m}$ to $2000~\mathrm{\mu m}$ and spans $10$ data points. 

   Due to the stochastic nature of the clump distribution, there exists an additional dependence of the spectral index on the locations of the clumps. To quantify this influence and to consider it in the analysis of our simulations, we repeat each simulation 10 times with different seeds for the random generator. The results with different random seeds are averaged and provide a mean value and a standard deviation. The standard deviation is in the order of $\sigma_{\alpha_\mathrm{mm}}\sim0.01$. Combined with the uncertainties of the simulation and the fitting process, we assume an effective error of $\sigma_\mathrm{eff}\sim0.02$.

   \begin{figure*}[t]
      \begin{subfigure}[b]{0.48\textwidth}
	  \includegraphics[width=\hsize]{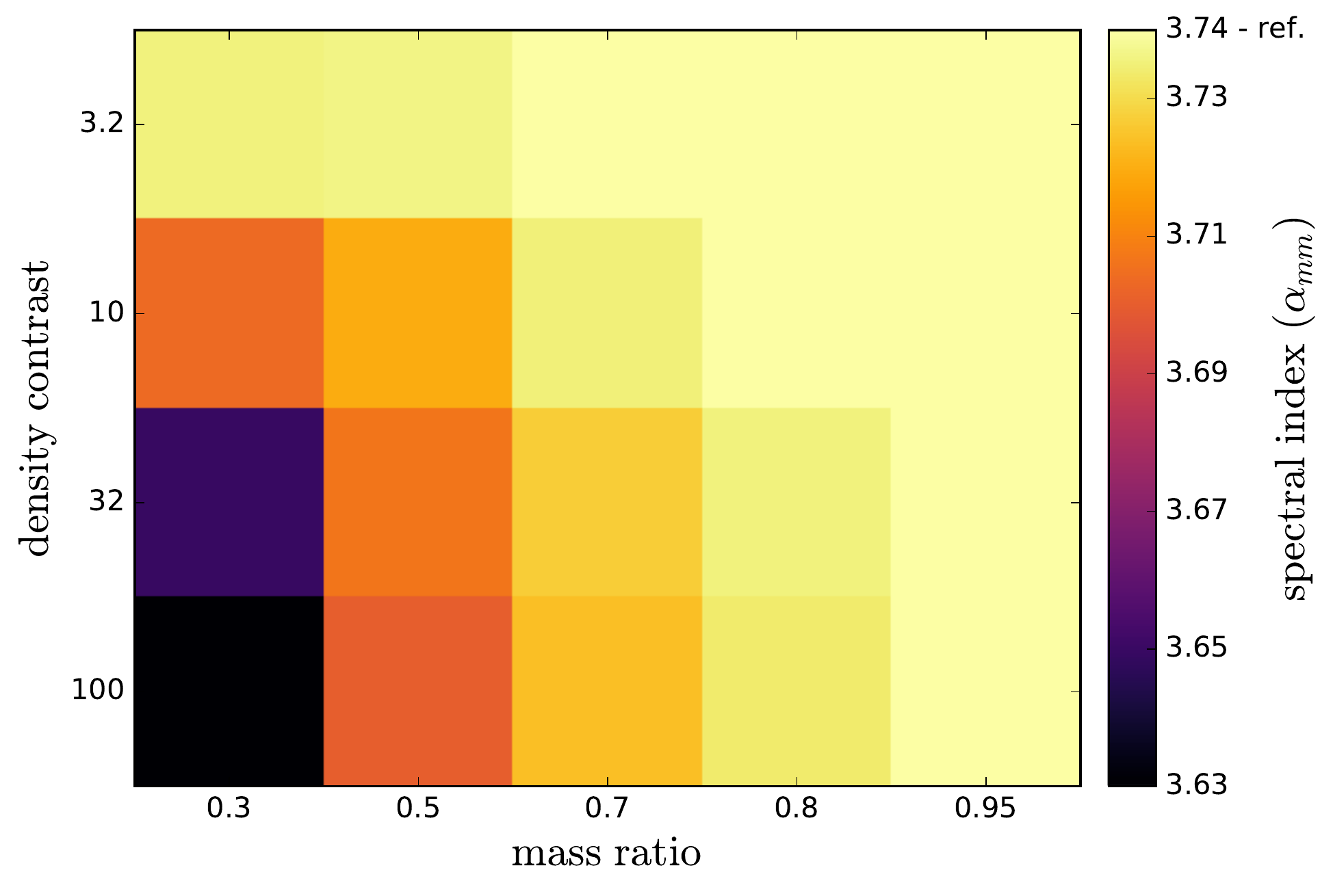}
	  \caption{$i=0^\circ$, $M_\mathrm{dust}=10^{-6}~\mathrm{M_\odot}$, $a_\mathrm{min}=0.005~\mathrm{\mu m}$ and $a_\mathrm{max}=0.25~\mathrm{\mu m}$.}
	  \label{fig:plot_mm_slope_E6_e}
      \end{subfigure}
      \begin{subfigure}[b]{0.48\textwidth}
	  \includegraphics[width=\hsize]{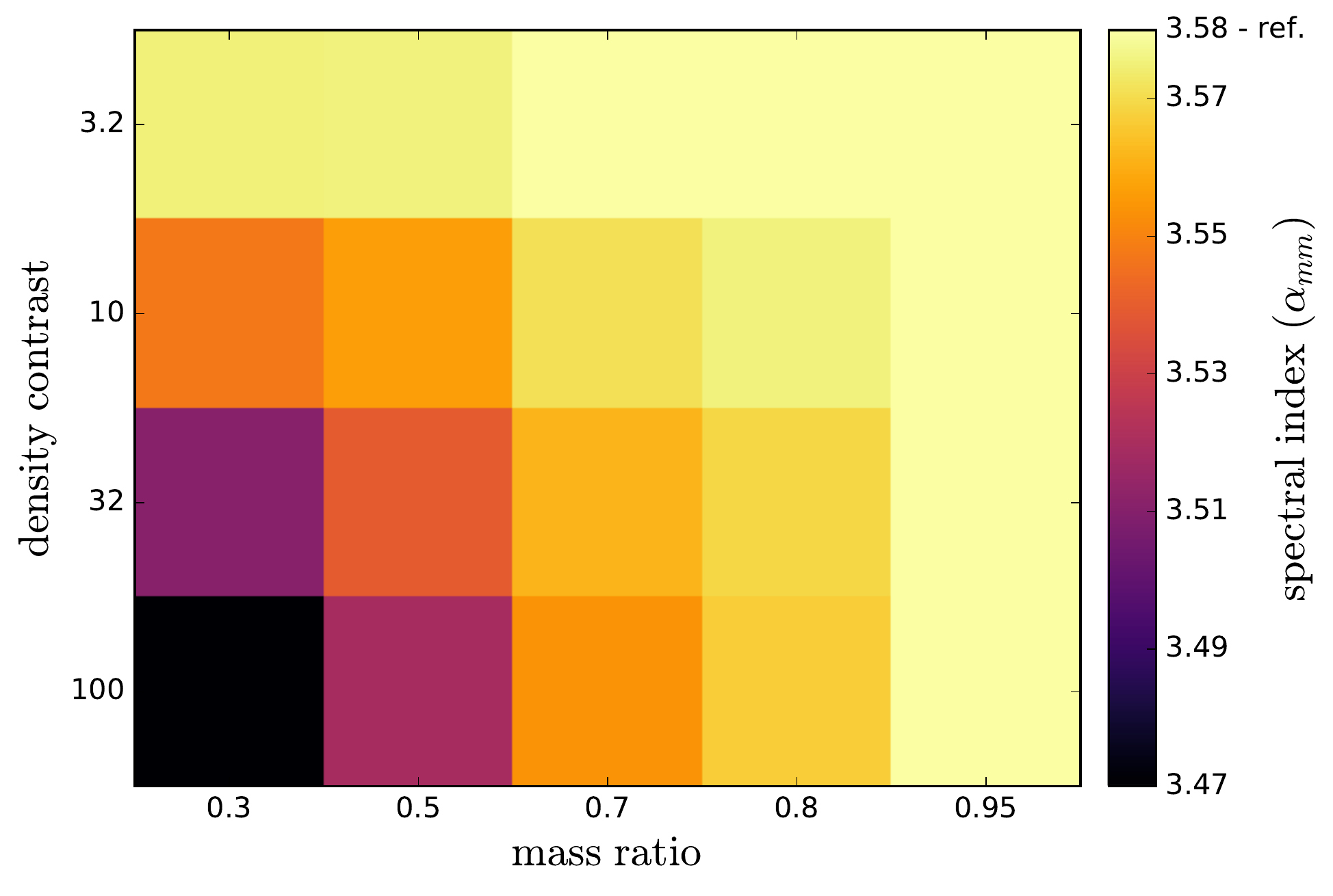}
	  \caption{$i=0^\circ$, $M_\mathrm{dust}=10^{-5}~\mathrm{M_\odot}$, $a_\mathrm{min}=0.005~\mathrm{\mu m}$ and $a_\mathrm{max}=0.25~\mathrm{\mu m}$.}
	  \label{fig:plot_mm_slope_E5_e}
      \end{subfigure}\\
      \begin{subfigure}[b]{0.48\textwidth}
	  \includegraphics[width=\hsize]{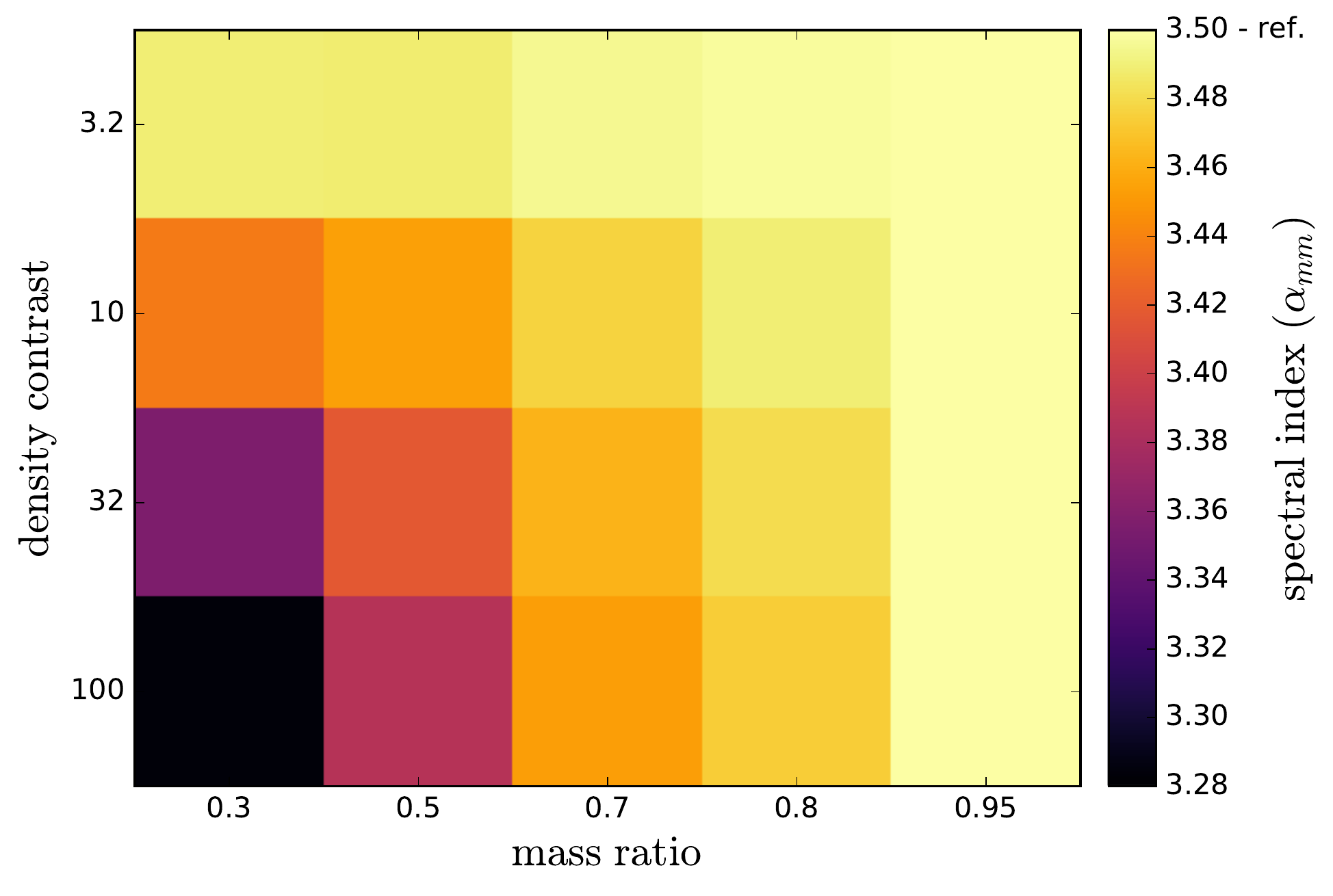}
	  \caption{$i=0^\circ$, $M_\mathrm{dust}=10^{-4}~\mathrm{M_\odot}$, $a_\mathrm{min}=0.005~\mathrm{\mu m}$ and $a_\mathrm{max}=0.25~\mathrm{\mu m}$.}
	  \label{fig:plot_mm_slope_E4_e}
      \end{subfigure}
      \begin{subfigure}[b]{0.48\textwidth}
	  \includegraphics[width=\hsize]{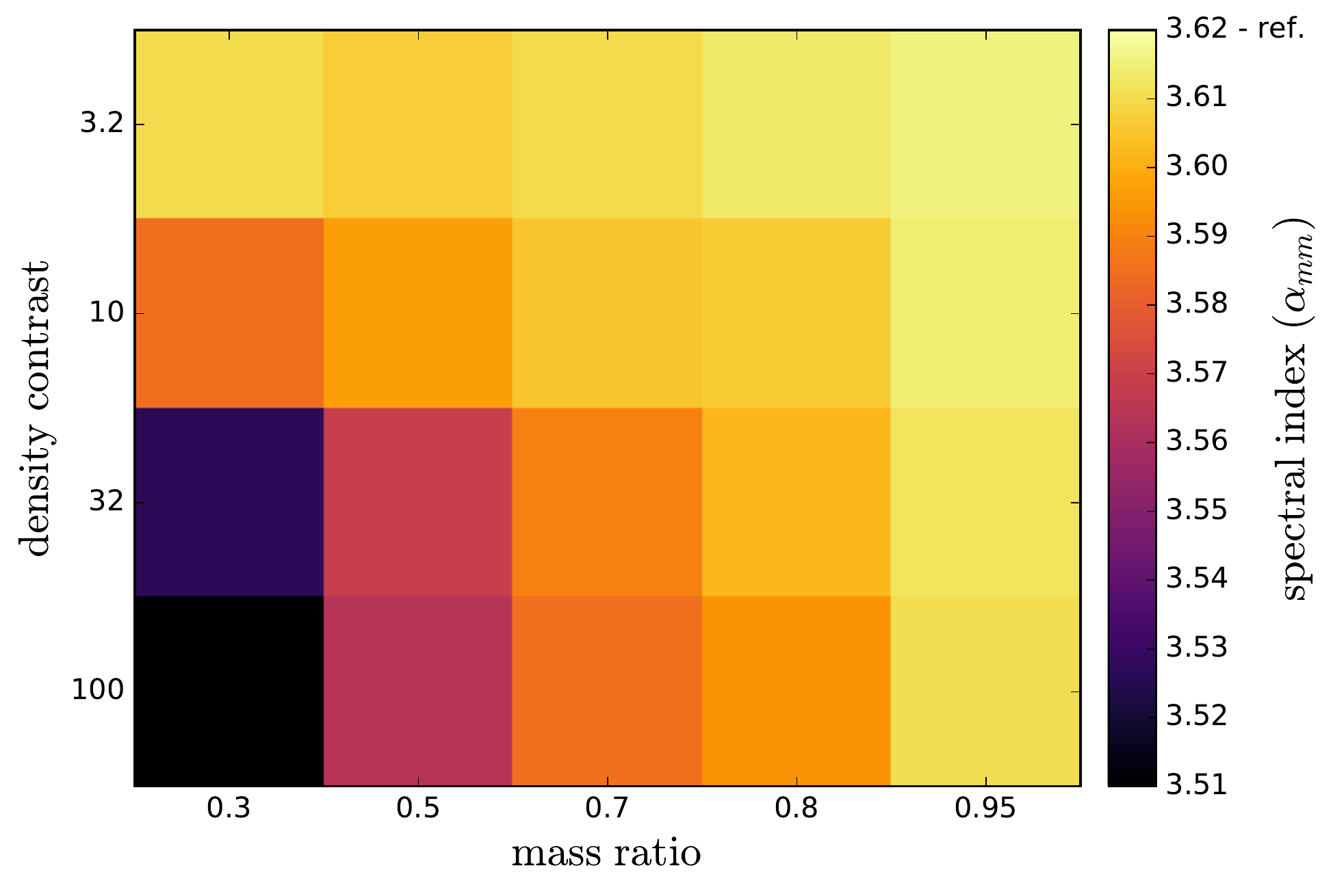}
	  \caption{$i=60^\circ$, $M_\mathrm{dust}=10^{-6}~\mathrm{M_\odot}$, $a_\mathrm{min}=0.005~\mathrm{\mu m}$ and $a_\mathrm{max}=0.25~\mathrm{\mu m}$.}
	  \label{fig:plot_mm_slope_K6_e}
      \end{subfigure}\\
      \begin{subfigure}[b]{0.48\textwidth}
	  \includegraphics[width=\hsize]{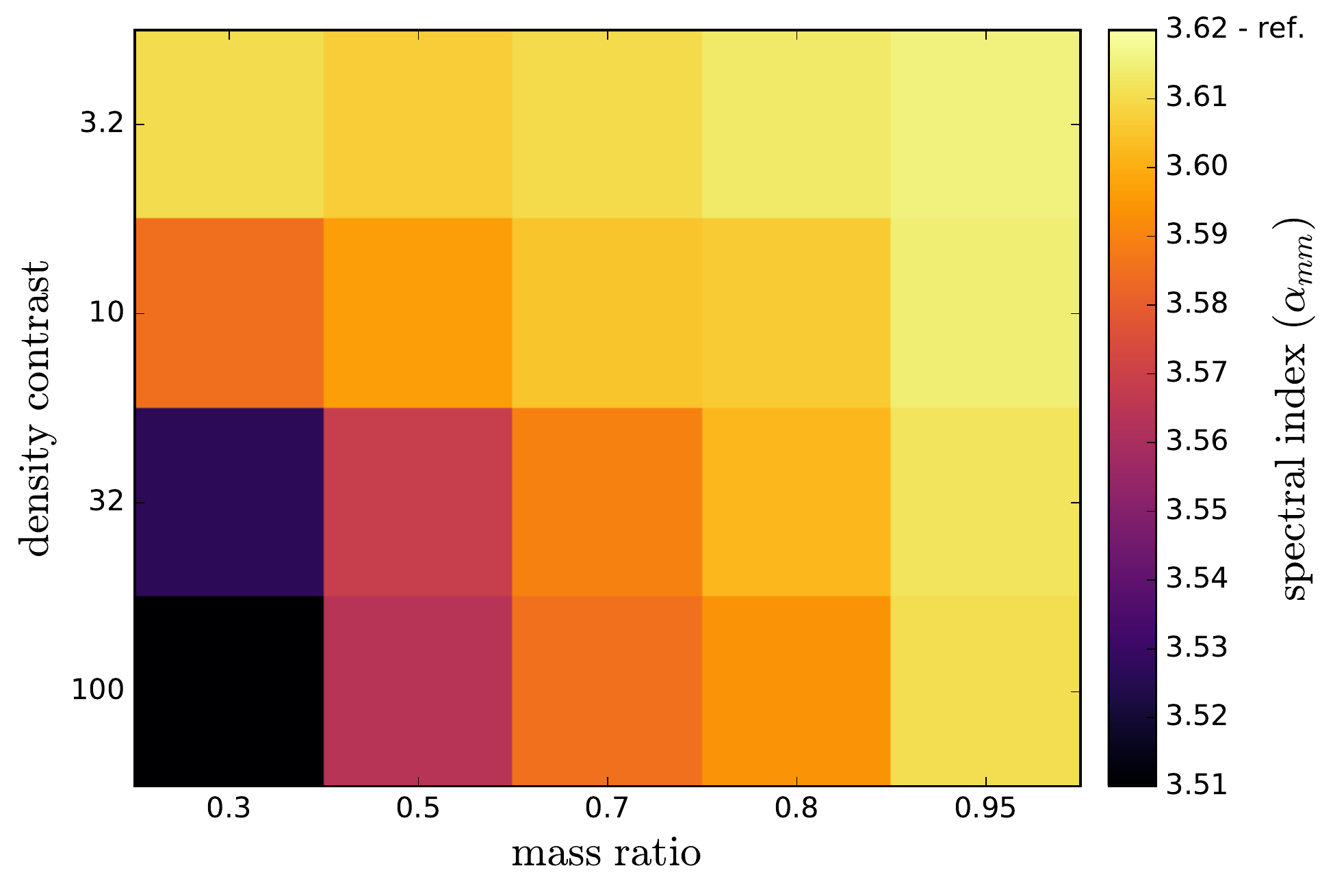}
	  \caption{$i=90^\circ$, $M_\mathrm{dust}=10^{-6}~\mathrm{M_\odot}$, $a_\mathrm{min}=0.005~\mathrm{\mu m}$ and $a_\mathrm{max}=0.25~\mathrm{\mu m}$.}
	  \label{fig:plot_mm_slope_J6_e}
      \end{subfigure}
      \begin{subfigure}[b]{0.48\textwidth}
	  \includegraphics[width=\hsize]{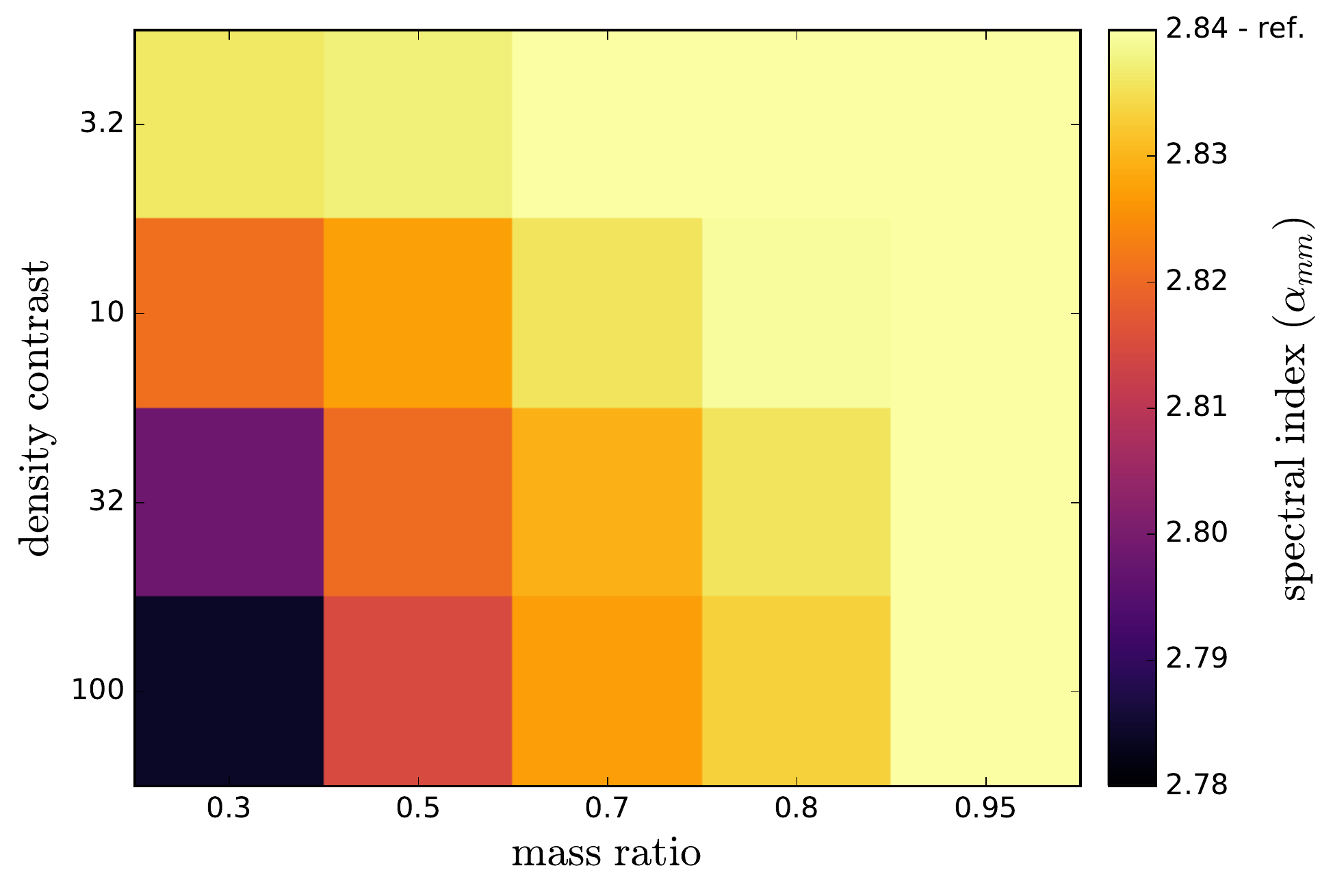}
	  \caption{$i=0^\circ$, $M_\mathrm{dust}=10^{-6}~\mathrm{M_\odot}$, $a_\mathrm{min}=0.005~\mathrm{\mu m}$ and $a_\mathrm{max}=10.0~\mathrm{\mu m}$.}
	  \label{fig:plot_mm_slope_H6_e}
      \end{subfigure}
      \caption{Spectral index of selected clumpy disks as a function of mass ratio and density contrast. The spectral index of the reference disk is marked with \textit{- ref}. The inclination, the dust mass and the dust grain size distribution is stated under each figure.}
   \end{figure*}

   As illustrated in Figs. \ref{fig:plot_mm_slope_E6_e} to \ref{fig:plot_mm_slope_H6_e}, each clumpy disk exhibits a lower spectral index than the related reference disk as expected from Eq. \ref{eqn:spectral_index_third}. Furthermore, the spectral index features almost the same pattern with mass ratio and density contrast for every used inclination, dust mass and dust grain size distribution. The clumpy density distribution with the highest density contrast and lowest mass ratio causes the strongest decrease in the spectral index. This parameter set produces also the lowest number of clumps that are the densest ones (see Fig. \ref{fig:density_dist}). Similar to the density distribution, disks with low density contrast and high mass ratio cause almost the same spectral index as the reference disk. From our simulations, we can provide a maximum decrease in the spectral index which is $\Delta\alpha_\mathrm{mm}\sim0.22$, corresponding to $\sim6\%$ of the spectral index of the reference disk.

   An increase in the disk mass results in a general decrease in the spectral index and a change of its range of possible values (see Figs. \ref{fig:plot_mm_slope_E6_e}, \ref{fig:plot_mm_slope_E5_e}, \ref{fig:plot_mm_slope_E4_e}, and Tabl. \ref{tab:E6_E5_E4}). The decrease in the spectral index is caused by the increased mass of the disk, resulting in an increased fraction of optically thick regions that reduce the amount of dust grains that contribute to the net flux. Considering the uncertainties of our simulations, the range of the spectral index broadens with increasing dust mass. This is caused by the increased optical depths of the clumps that are mainly responsible for the decrease in the spectral index.

   {\renewcommand{\arraystretch}{1.2}
   \begin{table}[h]
   \caption{Behavior of the spectral index of clumpy disks with selected dust masses. (see Figs. \ref{fig:plot_mm_slope_E6_e}, \ref{fig:plot_mm_slope_E5_e} and \ref{fig:plot_mm_slope_E4_e})}
   \label{tab:E6_E5_E4}
   \begin{tabular}{cccc}
    \hline
    \hline
     $M_\mathrm{dust}$ & $\alpha_\mathrm{mm}^\mathrm{ref}$ & $\alpha_\mathrm{mm}^\mathrm{min}$ & $(\alpha_\mathrm{mm}^\mathrm{ref}-\alpha_\mathrm{mm}^\mathrm{min})$ \\
    \hline
     $10^{-6}~\mathrm{M_\odot}$ & $3.74$ & $3.63$ & $0.11$ \\
     $10^{-5}~\mathrm{M_\odot}$ & $3.58$ & $3.47$ & $0.11$ \\
     $10^{-4}~\mathrm{M_\odot}$ & $3.50$ & $3.28$ & $0.22$ \\
    \hline
   \end{tabular}
   \end{table}
   {\renewcommand{\arraystretch}{1}

   Our increase in the inclination causes a general decrease in the spectral index, but its range of possible values remains constant (see Figs. \ref{fig:plot_mm_slope_E6_e}, \ref{fig:plot_mm_slope_K6_e}, \ref{fig:plot_mm_slope_J6_e}, and Tabl. \ref{tab:E6_K6_J6}). If a disk is seen at high inclination, the column density increases which decreases the spectral index. The constant range indicates that the impact of clumpy density distributions on the spectral index is independent on the disk inclination.
  
   {\renewcommand{\arraystretch}{1.2}
   \begin{table}[h]
   \caption{Behavior of the spectral index of clumpy disks with selected disk inclinations. (see Figs. \ref{fig:plot_mm_slope_E6_e}, \ref{fig:plot_mm_slope_K6_e} and \ref{fig:plot_mm_slope_J6_e})}
   \label{tab:E6_K6_J6}
   \begin{tabular}{cccc}
    \hline
    \hline
     $i$ & $\alpha_\mathrm{mm}^\mathrm{ref}$ & $\alpha_\mathrm{mm}^\mathrm{min}$ & $(\alpha_\mathrm{mm}^\mathrm{ref}-\alpha_\mathrm{mm}^\mathrm{min})$ \\
    \hline
     $0^\circ$ & $3.74$ & $3.63$ & $0.11$ \\
     $60^\circ$ & $3.71$ & $3.60$ & $0.11$ \\
     $90^\circ$ & $3.62$ & $3.51$ & $0.11$ \\
    \hline
   \end{tabular}
   \end{table}
   {\renewcommand{\arraystretch}{1}

   An increase in the dust grain size causes a decrease in the spectral index as expected from the wavelength dependence of the absorption efficiency $Q_\mathrm{abs}(\lambda)$ (see Figs. \ref{fig:plot_mm_slope_E6_e}, \ref{fig:plot_mm_slope_H6_e} and Tabl. \ref{tab:E6_H6}). In addition to the dust emissivity, optically thick regions in the density distribution feature a weaker decrease with wavelength due to the flatter $Q_\mathrm{abs}(\lambda)$ of larger grains. Consequently, the impact of optically thick regions (clumps) on the spectral index decreases.

   {\renewcommand{\arraystretch}{1.2}
   \begin{table}[h]
   \caption{Behavior of the spectral index of clumpy disks with selected maximum dust grain sizes. (see Figs. \ref{fig:plot_mm_slope_E6_e} and \ref{fig:plot_mm_slope_H6_e})}
   \label{tab:E6_H6}
   \begin{tabular}{cccc}
    \hline
    \hline
     $a_\mathrm{max}$ & $\alpha_\mathrm{mm}^\mathrm{ref}$ & $\alpha_\mathrm{mm}^\mathrm{min}$ & $(\alpha_\mathrm{mm}^\mathrm{ref}-\alpha_\mathrm{mm}^\mathrm{min})$ \\
    \hline
     $0.25~\mathrm{\mu m}$ & $3.74$ & $3.63$ & $0.11$ \\
     $10.0~\mathrm{\mu m}$ & $2.84$ & $2.78$ & $0.06$ \\
    \hline
   \end{tabular}
   \end{table}
   {\renewcommand{\arraystretch}{1}

%%%%%%%%%%%%%%%%%%%%%%%%%%%%%%%%%%%%%%%%%%%%%%%%%%%%%%%%
\subsection{Radial brightness profile}\label{r_b_p}
%%%%%%%%%%%%%%%%%%%%%%%%%%%%%%%%%%%%%%%%%%%%%%%%%%%%%%%%

   \begin{figure*}[ht]
    \resizebox{\hsize}{!}{\includegraphics[width=\hsize]{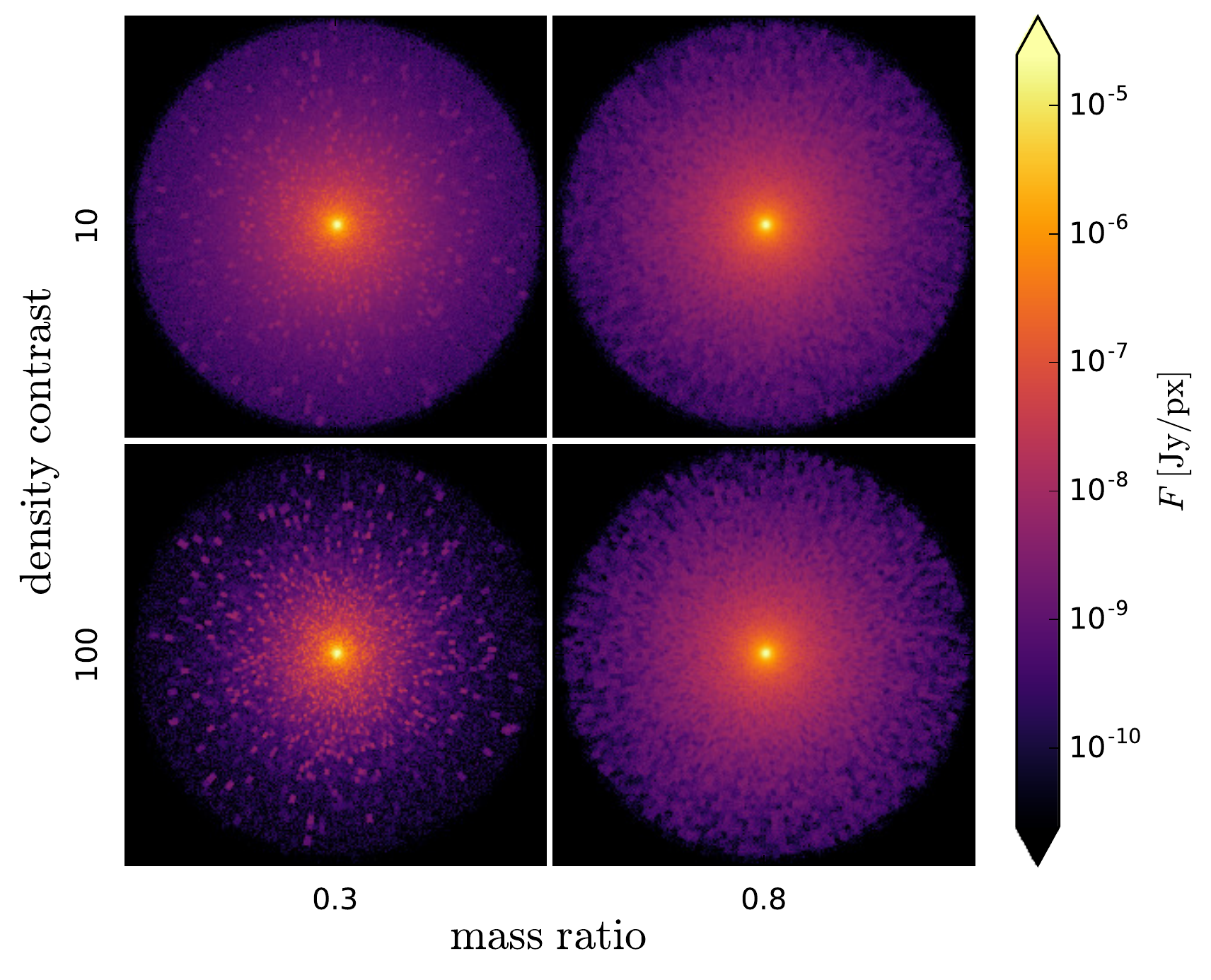}}
    \caption{Scattered light images dependent on mass ratio and density contrast. ($\lambda=0.726~\mathrm{\mu m}$, $i=0^\circ$, $M_\mathrm{dust}=10^{-6}~\mathrm{M_\odot}$, $a_\mathrm{min}=0.005~\mathrm{\mu m}$ and $a_\mathrm{max}=0.25~\mathrm{\mu m}$)}
    \label{fig:scattering_image_e}
   \end{figure*}

   We now investigate whether the density structure of circumstellar disks can be derived from the structure in scattered light images. For this purpose, we simulated scattered light images of a circumstellar disk for each combination of mass ratio and density contrast. Disks with four selected parameter settings are illustrated in Fig. \ref{fig:scattering_image_e}. All disks are seen face-on and have a dust mass of $10^{-6}~\mathrm{M_\odot}$. Their dust grain size distribution extends from $0.005~\mathrm{\mu m}$ to $0.25~\mathrm{\mu m}$. The scattered light images are simulated at $0.7~\mathrm{\mu m}$.

   To pursue our aim, we create a radial brightness profile from the scattered light images of the reference and two clumpy disks. The two clumpy disks have either more ($\mathit{\eta}=0.3$, $\mathit{k}=100$) or less ($\mathit{\eta}=0.8$, $\mathit{k}=100$) dense clumps. Each brightness profile is calculated by dividing the image in 50 equidistant concentric rings and averaging the flux of each ring. A standard deviation can be calculated from the variations in the flux in each of these rings. The standard deviation is not only influenced by brightness variations in azimuthal direction, but also in the radial direction of the equidistant concentric rings. Hence, the standard deviation of the reference disk is not zero. Because of this systematic error, the standard deviation of clumpy disks provides the impact of clumpy density distributions on the structure of scattered light images only in relation to the reference disk.

   \begin{figure*}[ht]
    \resizebox{\hsize}{!}{\includegraphics[width=\hsize]{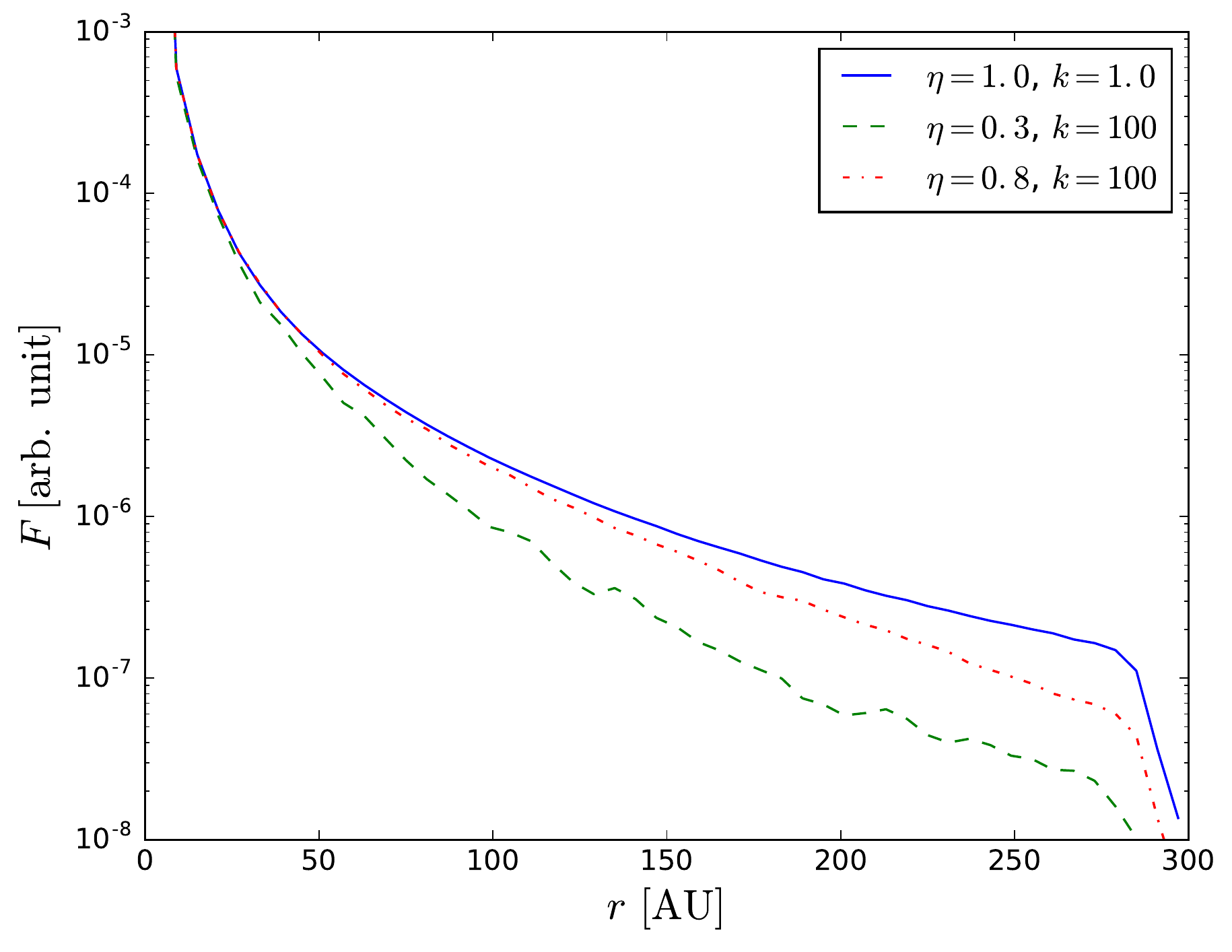} \qquad \includegraphics[width=\hsize]{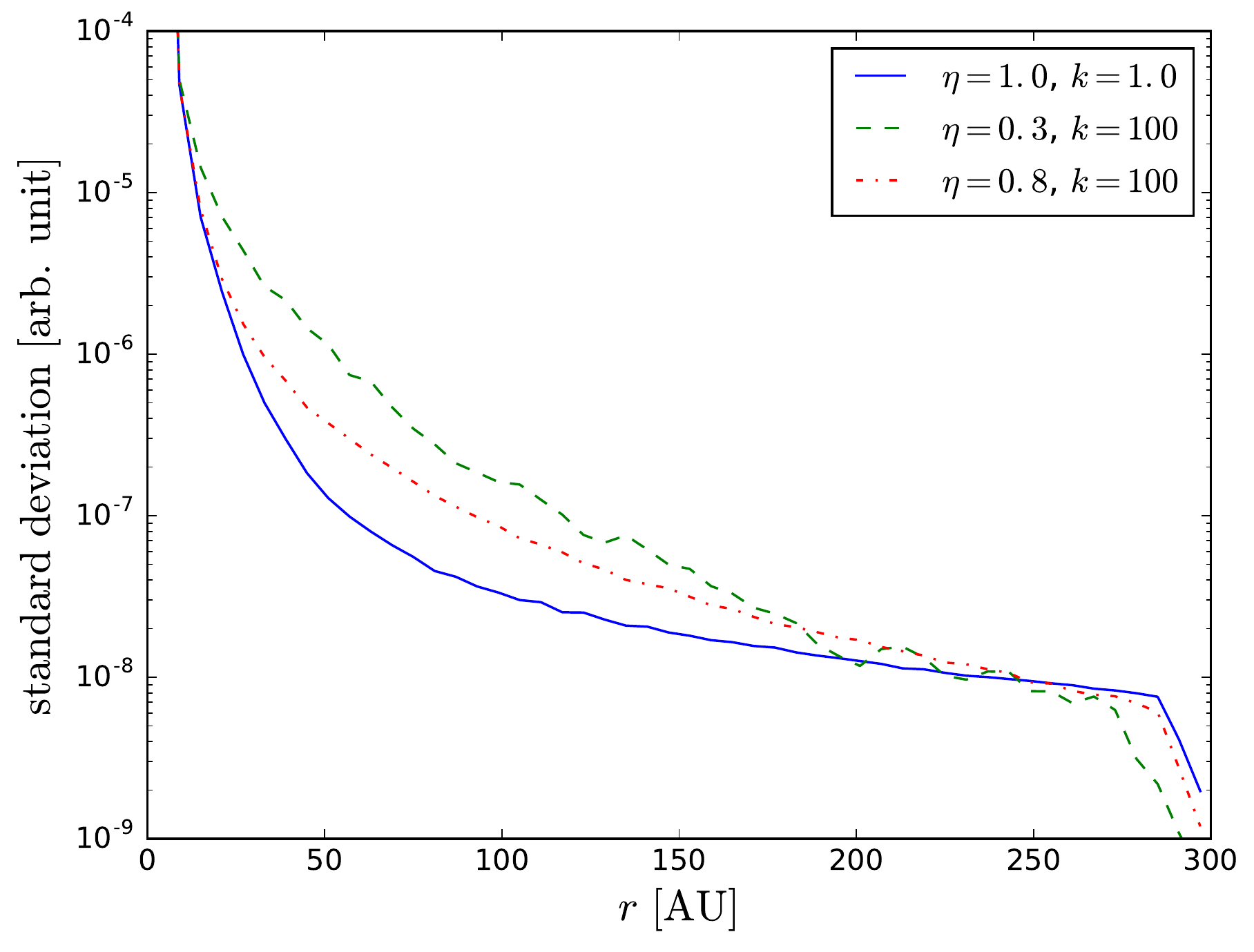}}
    \caption{Radial brightness profile (left) and its standard deviation (right) of the reference and two clumpy circumstellar disks. 50 concentric rings are used for the calculation of the brightness profile. ($\lambda=0.726~\mathrm{\mu m}$, $i=0^\circ$, $M_\mathrm{dust}=10^{-6}~\mathrm{M_\odot}$, $a_\mathrm{min}=0.005~\mathrm{\mu m}$ and $a_\mathrm{max}=0.25~\mathrm{\mu m}$)}
    \label{fig:rbp_clumpy_flux_e}
   \end{figure*}

   Figure \ref{fig:rbp_clumpy_flux_e} (left) illustrates that the reference disk exhibits higher fluxes than clumpy disks at almost every distance to the star. This behavior can be related to the shadowing effect of the clumps (if optically thick). The blocking of radiation also explains why the brightness profile becomes steeper with denser clumps. %(see Tabl. \ref{tab:rbp}). 
   The brightness profile of the density distribution with $\mathit{\eta}=0.3$ fluctuates with distance to the star which can be related to the stochastic distribution of the very dense clumps.

   The standard deviation features a different behavior than the brightness profile (see Fig. \ref{fig:rbp_clumpy_flux_e}, right). Clumpy disks show azimuthal flux variations that are proportional to the density of the clumps. In the innermost part of the disks the strong radial decrease in radiation outshines variations in azimuthal direction. In the outer regions, the combination of the shadowing effect and the high scattering probability of clumps results in a standard deviation that is almost the same for each density distribution. Therefore, the density structure can be inferred from the standard deviation in the radial brightness profile, but only in the range between $20~\mathrm{au}$ and $150~\mathrm{au}$. 

%%%%%%%%%%%%%%%%%%%%%%%%%%%%%%%%%%%%%%%%%%%%%%%%%%%%%%%%
\subsection{Polarization}\label{polarization}
%%%%%%%%%%%%%%%%%%%%%%%%%%%%%%%%%%%%%%%%%%%%%%%%%%%%%%%%

   \begin{figure*}
    \resizebox{\hsize}{!}{\includegraphics[width=\hsize]{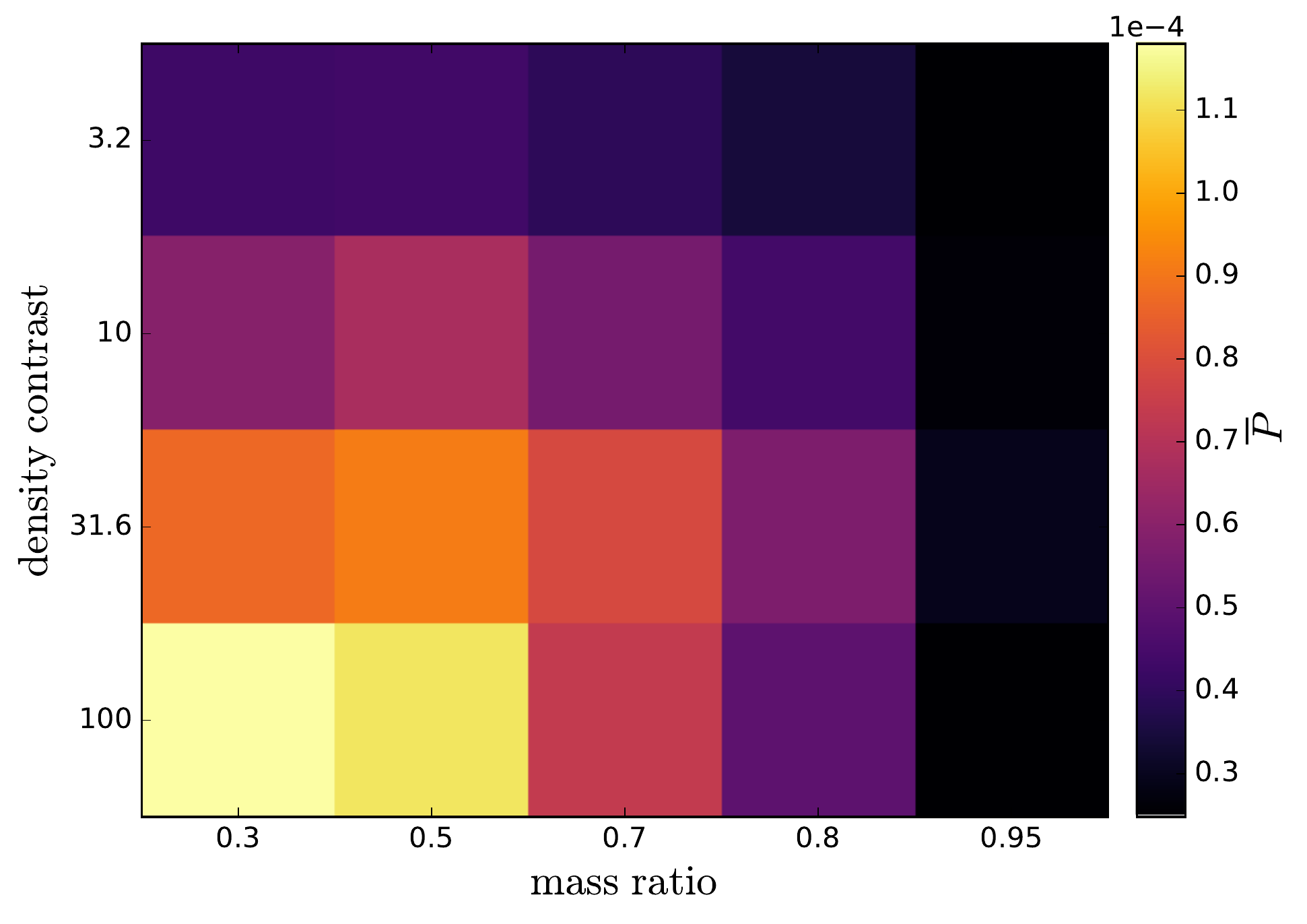} \qquad \includegraphics[width=\hsize]{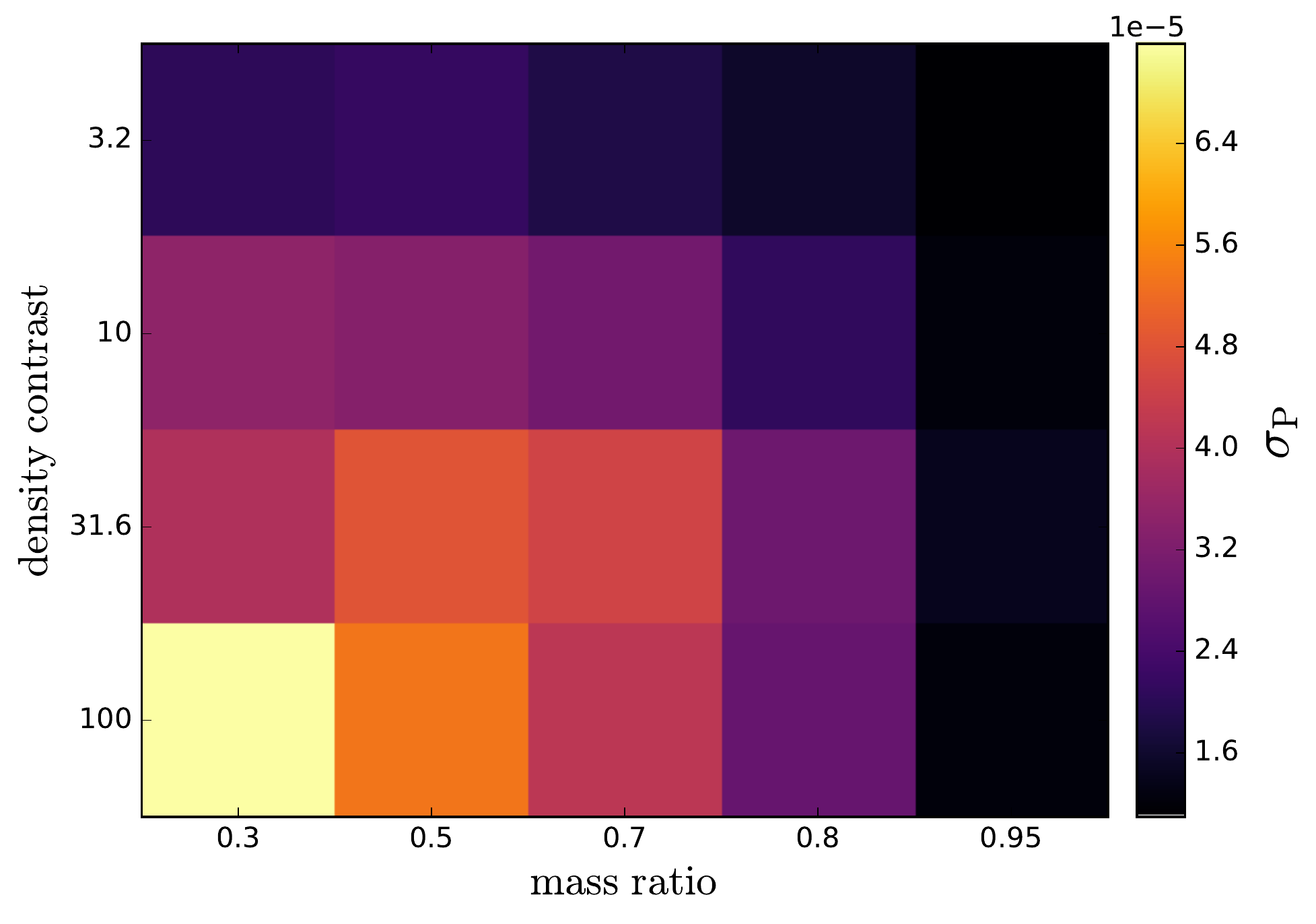}}
    \caption{Mean degree of polarization (left) and its standard deviation (right) dependent on mass ratio and density contrast. Each configuration is simulated with 100 different positions of clumps. ($\lambda=0.726~\mathrm{\mu m}$, $i=0^\circ$, $M_\mathrm{dust}=10^{-6}~\mathrm{M_\odot}$, $a_\mathrm{min}=0.005~\mathrm{\mu m}$ and $a_\mathrm{max}=0.25~\mathrm{\mu m}$)}
    \label{fig:net_pol_mean}
   \end{figure*}

   In the previous section, we showed that clumps have a great impact on the scattered light of circumstellar disks. It stands to reason that clumps also have a significant impact on the degree of polarization of the scattered light. If this is true, the estimation of the density structure with the degree of polarization might be achievable. To investigate this, we simulate the scattered light of clumpy circumstellar disks with $\lambda=0.7~\mathrm{\mu m}$, $i=0^\circ$, $M_\mathrm{dust}=10^{-6}~\mathrm{M_\odot}$, $a_\mathrm{min}=0.005~\mathrm{\mu m}$ and $a_\mathrm{max}=0.25~\mathrm{\mu m}$ for each combination of mass factor and density contrast. Subsequently, we calculate the degree of polarization $P$ of each scattered light image according to the following equation:
   \begin{equation}
     P=\frac{\sqrt{F_\mathrm{Q}^2+F_\mathrm{U}^2}}{F_\mathrm{I}}.
   \end{equation}
   Here, $F_\mathrm{Q}$ and $F_\mathrm{U}$ are the linear polarization components of the Stokes vector where $F_\mathrm{I}$ is the intensity component. The circular component $F_\mathrm{V}$ is negligible compared to the linear components and therefore not considered.

   In contrast to the spectral index, the polarization exhibits very strong variations with different positions of the clumps. To consider this, we simulate the circumstellar disks 100 times for each set of parameters with different random seeds in our clump distribution algorithm (see Sect. \ref{clumpy_distr}). We calculate the mean degree of polarization and the standard deviation for each set of parameters. 
   
   As illustrated in Fig. \ref{fig:net_pol_mean}, the degree of polarization and its standard deviation show a clear increase with denser clumps which looks similar to the pattern of Figs. \ref{fig:plot_mm_slope_E6_e} to \ref{fig:plot_mm_slope_H6_e}. The clumps have higher and the interclump medium lower scattering probabilities than corresponding regions in the reference disk which cause a highly on the clump position dependent degree of polarization. For illustration, Fig. \ref{fig:scattering_Q} shows the $F_\mathrm{Q}$ component of the Stokes vector of the clumpy disk with $\mathit{\eta}=0.3$ and $k=32$. One can clearly see that the clumps disturb the symmetry of the disk and, in relation to their number, cause an increase in the degree of polarization. Additionally, the standard deviation amounts to $50\%$ of the corresponding degree of polarization so that a change in the clump locations can change the polarization in the same way as a change of mass ratio or density contrast. As a result, a given degree of polarization can be produced by different clumpy density distributions and clump positions.

   \begin{figure}
    \resizebox{\hsize}{!}{\includegraphics[width=\hsize]{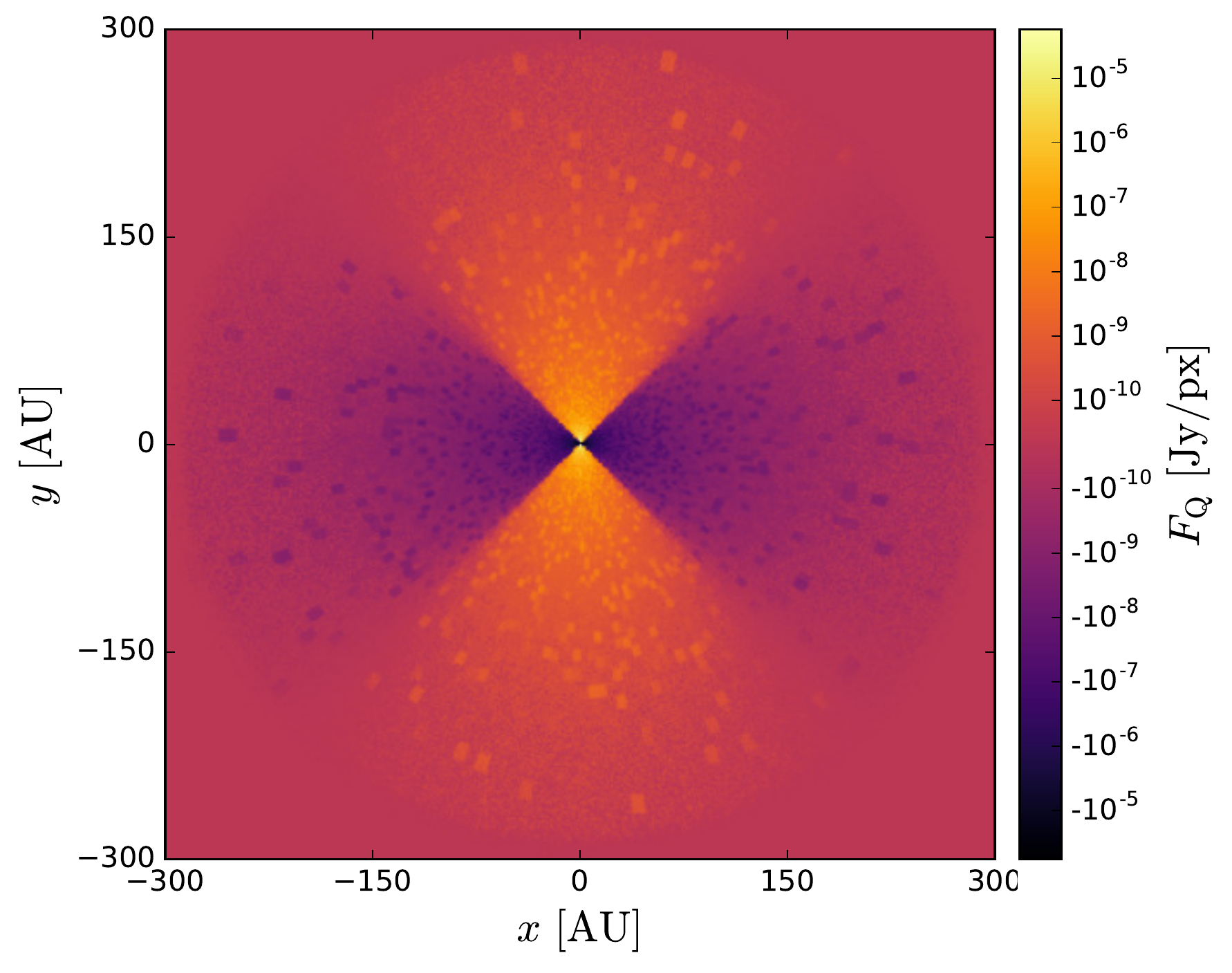}}
    \caption{$F_\mathrm{Q}$-component of the Stokes vector of a clumpy disk with $\mathit{\eta}=0.3$ and $k=32$. \\ ($\lambda=0.726~\mathrm{\mu m}$, $i=0^\circ$, $M_\mathrm{dust}=10^{-6}~\mathrm{M_\odot}$, $a_\mathrm{min}=0.005~\mathrm{\mu m}$ and $a_\mathrm{max}=0.25~\mathrm{\mu m}$)}
    \label{fig:scattering_Q}
   \end{figure}

   \begin{figure*}
    \resizebox{\hsize}{!}{\includegraphics[width=\hsize]{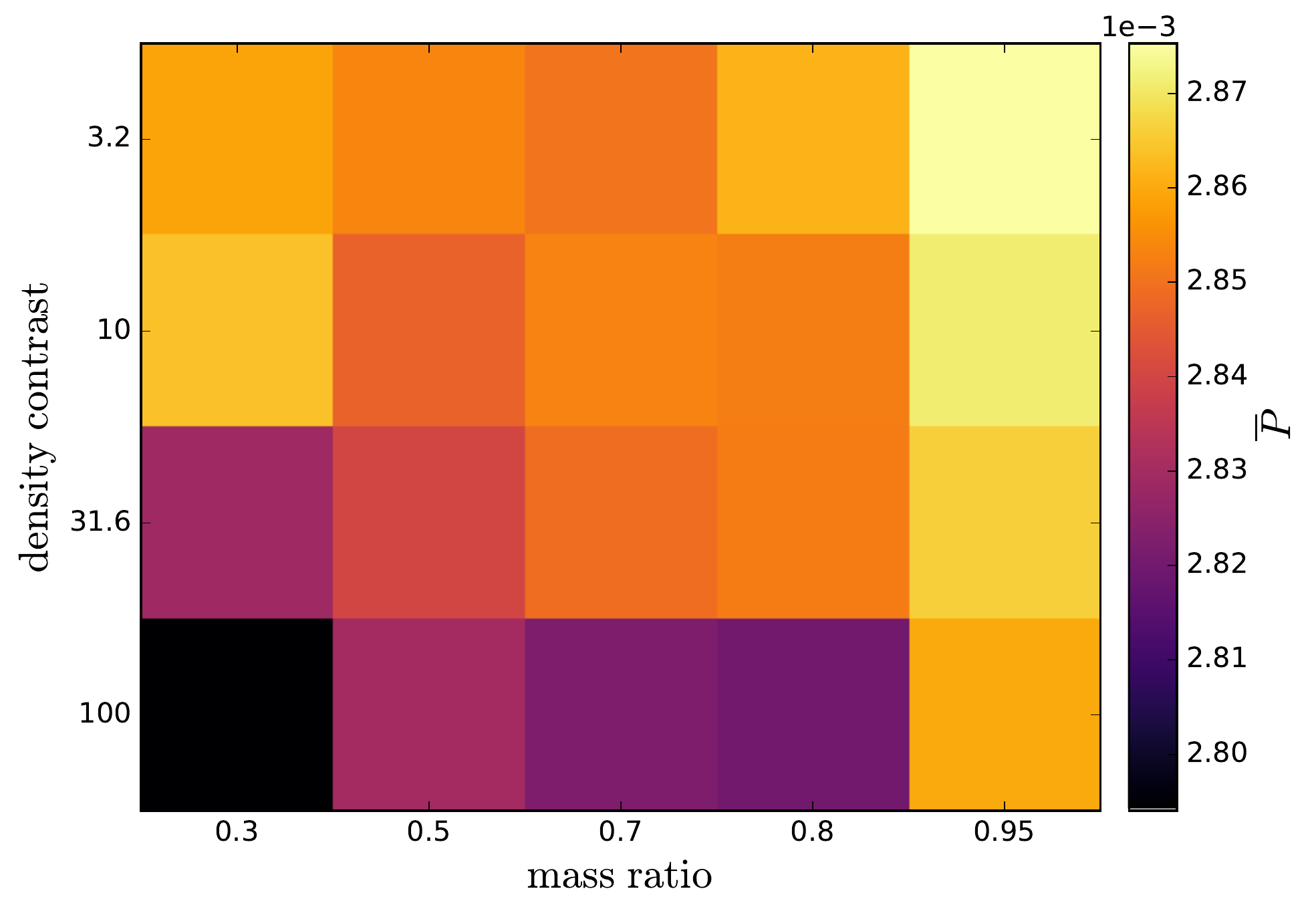} \qquad \includegraphics[width=\hsize]{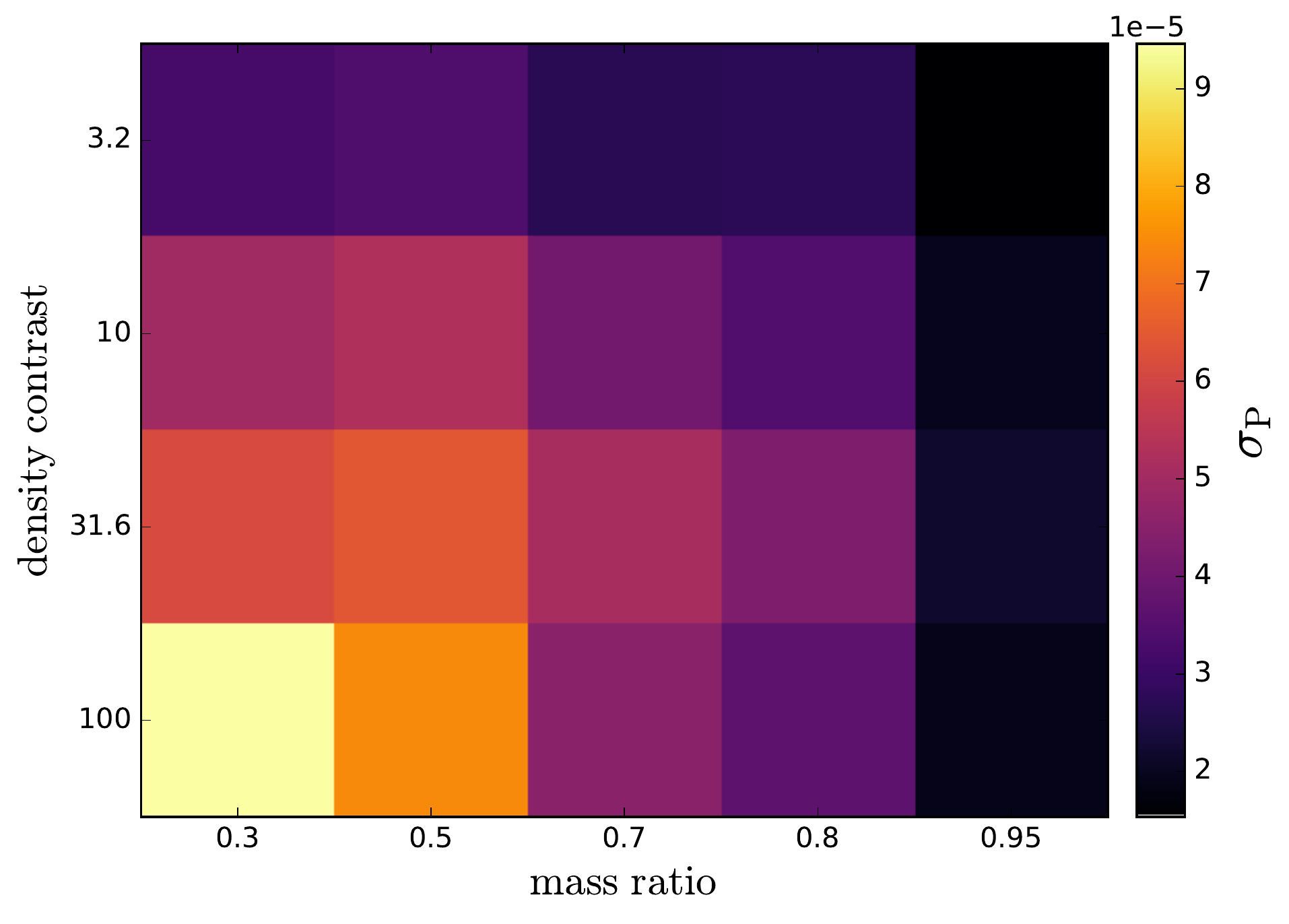}}
    \caption{Mean degree of polarization (left) and its standard deviation (right) dependent on mass ratio and density contrast. Each configuration is simulated with 100 different positions of clumps. ($\lambda=0.726~\mathrm{\mu m}$, $i=45^\circ$, $M_\mathrm{dust}=10^{-6}~\mathrm{M_\odot}$, $a_\mathrm{min}=0.005~\mathrm{\mu m}$ and $a_\mathrm{max}=0.25~\mathrm{\mu m}$)}
    \label{fig:net_pol_mean_A}
   \end{figure*}

   Our results may change in case of disks with an inclination $\neq0$. Because of the preferred direction, the $F_\mathrm{Q}$ and/or $F_\mathrm{U}$ component of the Stokes vector has more regions with positive (negative) values than with negative (positive). This causes a degree of polarization $\neq0$. Therefore, we repeat the simulations above with an inclination of $45^\circ$ (see. Fig. \ref{fig:net_pol_mean_A}). Each disk with a clumpy density distribution features a lower degree of polarization than the reference disk. As mentioned in Sect. \ref{r_b_p}, the shadowing effect of the clumps reduces the observed scattered light. In combination with the preferred direction, the positive (negative) flux of the $F_\mathrm{Q}$ and/or $F_\mathrm{U}$ component is more reduced than the negative (positive). As a consequence, the degree of polarization decreases with denser clumps. The standard deviation in the degree of polarization shows the same trend as the face-on disks in Fig. \ref{fig:net_pol_mean}. Furthermore, it is in the same order of magnitude as the deviation between the degree of polarization of the clumpy disks and the related reference disk. Hence, a given degree of polarization can be produced by different clumpy density distributions and clump positions. Furthermore, simulations with other inclinations $\neq0$ lead to the same results.

%%%%%%%%%%%%%%%%%%%%%%%%%%%%%%%%%%%%%%%%%%%%%%%%%%%%%%%%
\subsection{Clump length scales}\label{clump_size}
%%%%%%%%%%%%%%%%%%%%%%%%%%%%%%%%%%%%%%%%%%%%%%%%%%%%%%%%

Our results are applicable for clumpy density distributions with clump length scales from a few $0.01~\mathrm{au}$ in the innermost regions to some $10~\mathrm{au}$ in the outermost regions. However, simulations of clumpy disks with other length scales show a clear trend which is outlined in the following.

The influence of clumpy density distributions on the temperature distribution of a circumstellar disk is almost independent of the clump size as long as the clumps are  optically thick in the transition region and divide the midplane into optically thick and thin areas. Due to computational limitations, we are only able to decrease the clump length scales by a factor of three. Even at this size, the clumps are big enough to fulfill these requirements.

In clumpy density distributions with larger clumps it is more likely to have more line-of-sights with a high or low optical depth, whereby the number of line-of-sights with a medium optical depth decreases. The submm/mm spectral index depends on the amount and optical depth of those line-of-sights that are not optically thin. Consequently, the decrease in the spectral index depends directly on the length scales of the clumps.

Larger clumps in clumpy density distributions show a more locally concentrated impact on the scattered stellar radiation. This causes greater fluctuations in the radial brightness profiles, but their general behavior as well as the standard deviation remains almost the same. For smaller clumps, the opposite effect occurs. Due to their more locally concentrated impact, the larger clumps cause on average a greater net-polarization. However, the influence of the global parameters $k$ and $\eta$ compared to the influence of the clump locations remains the same.

\section{Discussion}\label{discussion}

\subsection{Temperature distribution}\label{discussion_t}

   \cite{2008A&A...482...67S} showed that the spatial temperature distribution in clumpy tori of Seyfert galaxies exhibits a broad range at a given radial distance to the central heating source. We achieved this result for circumstellar disks too. In addition, clumpy circumstellar disks in the considered parameter space have an increased mean temperature in their transition region from the optically thin upper disk layers to the disk interior of up to $12~\mathrm{K}$ depending on the disk mass. Consequently, the inhomogeneous structure of circumstellar disks has a potential impact on the chemical composition of the gas and dust phase (e.g. ice layers), preferentially in the optically thin/optically thick transition region (e.g. \citealt{2009A&A...495..881V}).

   We also found out that the radial distance of isothermal lines varies by several $10~\mathrm{au}$ in the interior of circumstellar disks due to clumpy density distributions. In particular, a deviation in the ice-line can greatly influence the evolution of circumstellar disks and especially the formation of planets. Dust grains with frozen out ice layers could accumulate more mass in a shorter time period than expected for dust grains without ice-layers \citep{2013A&A...552A.137R}.

\subsection{Spectral index}

   It is expected that optically thick regions in circumstellar disks cause a decrease in the spectral index in the submm/mm of the SED. We found out, that clumps in the density distribution of circumstellar disks confirm this expectation. In the considered model space, the maximum decrease is about $0.22$ or $6\%$ of the spectral index of the reference disk. Due to the decrease in the spectral index, the dust grain size derived from the submm/mm-slope of the SED may be overestimated, if the density structure is not taken into account. If the behavior of the spectral index with the dust grain size is known, the maximum deviation in the spectral index can be used to constrain the maximum dust grain size. 

   As discussed in Sect. \ref{spectral_index}, we found that there is no simple correlation between the spectral index and the dust grain size. Selected parameter sets and fitting methods may result in a spectral index that does not monotonically decrease with increasing dust grain size.

\subsection{Radial brightness profiles}

   The scattered light radial brightness profile of clumpy circumstellar disks exhibits a decrease by up to one magnitude compared to the related reference disk which is expected by the shadowing effect of the clumps (if optically thick). Therefore, smooth and continuous disks are more likely to be observed than clumpy disks. In addition, clumpy disks feature azimuthal brightness variations which are maximum in the medium radial extent of the disk. Therefore, we suggest to focus on this region for an observation of clumpy circumstellar disks in scattered light. The azimuthal brightness variations in the clumpy disks are only up to one order of magnitude larger than the radial brightness variations in the continuous disks.

\subsection{Polarization}\label{discussion_p}

   Clumpy density distributions disturb the symmetry of smooth and continuous circumstellar disks. Therefore, different values of our parameters $k$ and $\eta$ lead to different degrees of polarization. Nevertheless, an estimation of the disk structure with the degree of polarization of scattered light is not achievable, because the influence of the locations of the clumps is in the same order of magnitude as a change of $k$ and $\eta$.

\subsection{Limitations of the model}

   The underlying grid on which the density distribution is discretized (in our case: grid defined using spherical coordinates) has a strong influence on the shape and density of the clumps. Our clumps are getting bigger and less dense with increasing distance to the central star. In addition, our clump distribution algorithm features a small bias towards clumps closer to the central star. This assumptions are in contrast to the works by \cite{1996ApJ...463..681W} and \cite{1998A&A...340..103W}, but this kind of clumps should be more realistic for circumstellar disks than everywhere the same size, number and density. Furthermore, our results are in agreement with various expectations about clumpy circumstellar disks which also confirms our modeling approach.

   With the considered clump length scales, our results provide a characteristic overview of the influence of clumpy density distributions on physical and observational characteristics of circumstellar disks. Density distributions with larger clumps show our results even more pronounced whereby, with smaller clumps, the impact of clumpiness is less pronounced.

   Our studies rely on a two-phase medium representation of a clumpy disk that is in part complementary to previous approaches \citep{1996ApJ...463..681W, 1993MNRAS.264..145H, 2003MNRAS.342..453H, 2008A&A...482...67S, 2000ApJ...528..799W}. Given the limited angular resolution to derive and quantify the potential inhomogeneous structure of circumstellar disk, little constraints are currently available for the shape, density and distribution of clumps or other inhomogeneities. As a consequence, the best approach to implementing clumpy density distributions is still open. However, as summarized in Sect. \ref{discussion_t} - \ref{discussion_p}, the applied model allows derivation of general trends resulting from a clumpy/inhomogeneous density distribution of circumstellar disks.

\section{Conclusion}\label{conclusions}

    While circumstellar disks are usually modeled with a smooth and continuous density distribution, the structure is expected to differ strongly from continuous in reality. Therefore, we pursued two aims. On the one hand, we investigated how inhomogeneous density distributions change observational characteristics of circumstellar disks and their impact on selected internal physical and observational quantities. On the other hand, we investigated whether the clumpiness of circumstellar disks can be inferred from their physical and observable quantities. These characteristics are the temperature distribution, the spectral energy distribution, the radial brightness profile and the degree of polarization of scattered stellar radiation. Within the considered parameter space, we derived the following qualitative and quantitative conclusions for these quantities:
    
    \begin{enumerate}
       \item We found that the temperature in the interior of clumpy circumstellar disks varies by several Kelvin in azimuthal direction. Because of the low temperature gradient in the interior of circumstellar disks, the radial distance of isothermal lines differ by some $10~\mathrm{au}$ from the radial distance in continuous disks. For instance, dust grains with ice-layers in clumpy disks can exist closer to the star than in continuous disks. In addition, the average temperature in the transition region from optically thin to optically thick in clumpy disks is up to $12~\mathrm{K}$ lower than in continuous disks. This variation in temperature is expected to influence quantities related to the composition and aggregate state of the volatile species in the disk.

       \item Clumpy density distributions cause a decrease in the spectral index in the submm/mm range of the SED of circumstellar disks. At maximum this decrease is about $0.22$ or $6\%$ of the spectral index of the reference disk. The strength of this effect can be varied by changing the dust mass or grain size, but not by changing the inclination of the disk. As a consequence of the lower spectral index, the dust grain size derived from the submm/mm-slope of the SED may be overestimated, if the inhomogeneity of the disk density distribution is not taken into account.

       \item The radial brightness profile of clumpy circumstellar disks in scattered light exhibits a decrease by up to one order of magnitude compared to the related reference disk. In addition, clumpy disks feature azimuthally brightness variations in the scattered light regime which have their maximum at the medium radial extent of the disk. High-contrast observations of the structures of circumstellar disks in the optical should therefore focus on this region.

       \item Clumpy density distributions disturb the disk symmetry and therefore change the degree of polarization of scattered light in the optical. However, this effect depends both on the exact locations of the clumps and on the global clumpy parameter $k$ and $\mathit{\eta}$. For this reason it is impossible to derive unique constraints on the clumpiness of circumstellar disks from polarization observations in the scattered light.
    \end{enumerate}

   \noindent In conclusion, the structures of circumstellar disks are an important characteristic and must be considered to obtain adequate results.

\begin{acknowledgements}
      We wish to thank our colleague Peter Scicluna for his friendly assistance and helpful discussions. 
      Part of this work was supported by the German
      \emph{Deut\-sche For\-schungs\-ge\-mein\-schaft, DFG\/} project
      number WO 857/12-1.

\end{acknowledgements}

%-------------------------------------------------------------------
\bibliographystyle{aa}
\bibliography{bibtex}
%-------------------------------------------------------------------

\end{document}